%% file: main.tex
\documentclass[conference,10pt,a4paper]{IEEEtran}
\usepackage{graphicx} 
\usepackage{subfig}
\usepackage{float}
\usepackage{subfloat}
\usepackage{epstopdf}
\usepackage{color}
\usepackage{url} 
\usepackage{amsmath}

\usepackage{algorithm}
\usepackage[noend]{algpseudocode}

\makeatletter
\def\BState{\State\hskip-\ALG@thistlm}
\makeatother

\newif\ifconf
\newif\ifskip

\conftrue  
\conffalse 

\skipfalse

\ifconf
 \else
\fi

\newtheorem{prop}{Property}

\definecolor{blue}{rgb}{0,0,0}
\definecolor{red}{rgb}{0,0,0}

\begin{document}
\begin{sloppypar}
\ifconf
\else
\title{The Role of Inter-Controller Traffic \\ in the Placement of Distributed  SDN Controllers}
\fi

\ifconf
\else
\author{Tianzhu Zhang, Andrea~Bianco, Samuele De Domenico,
        Paolo~Giaccone  \\{\small Dept.\ Electronics and Telecommunications, Politecnico di Torino, Italy}}
\fi

\maketitle
\input{abstract}

\input{intro}

\input{controllers}
\input{validation}

\input{place}
\input{methodology}
\input{results}

\input{algorithm}

\input{previous}

\input{conclusion}
\bibliographystyle{IEEEtran}
\bibliography{IEEEabrv,mybiblio}


\end{sloppypar}
\end{document}

%% file: abstract.tex
\begin{abstract}
We consider a distributed Software Defined Networking (SDN) architecture 
adopting a cluster of multiple controllers to improve network performance 
and reliability. 
Besides the Openflow control traffic exchanged between controllers and switches, 
we focus on the control traffic exchanged among the controllers in the cluster,  
needed to run coordination and consensus algorithms to keep the controllers 
synchronized. We estimate the effect of the inter-controller communications 
on the reaction time perceived by the switches depending on the data-ownership model 
adopted in the cluster. 
\ifconf
\else
{\color{blue}The model is accurately validated in an operational  
Software Defined WAN (SDWAN). }
\fi
We advocate a careful placement of the controllers, that should take into account 
both the above kinds of control traffic. We evaluate, for some real ISP  
network topologies, the delay tradeoffs for the controllers 
\ifconf
\else
{\color{blue}
placement problem and we propose a novel evolutionary 
algorithm to find the corresponding Pareto frontier.}
\fi

Our work provides novel quantitative tools to optimize the planning 
and the design of the network supporting the control plane of SDN networks, 
especially when the network is very large and in-band control plane is adopted. 
We also show that for operational distributed controllers 
(e.g.\ OpenDaylight and ONOS), the location of the controller 
which acts as a leader in the consensus algorithm has a strong impact 
on the reactivity perceived by switches.
\end{abstract}


%% file: intro.tex

\section{Introduction}

The centralized network control of the Software Defined Networking (SDN) paradigm, 
which enables the development of complex network applications, 
poses two main issues. 
First, a limited reliability, due to the single point-of-failure. 
Second, the control traffic between the switches and the controller 
concentrates on a single server, whose processing capability is limited, 
creating scalability issues. 
Distributed SDN controllers are designed to address 
the above issues, while preserving a logically centralized view 
of the network state necessary to ease the development of network applications. 
In a distributed architecture, multiple controllers are responsible of the interaction 
with the switches, with two beneficial effects. First, the processing load 
at each controller decreases, because the control traffic between the switches 
and the controllers is distributed, with a beneficial load balancing effect. 
Second, resilience mechanisms are implemented to improve network reliability 
in case of controller failures. 

Distributed controllers adopt coordination protocols and algorithms to synchronize 
their shared data structures which define the network state and to enable 
a centralized view of the network state for the applications. 
The algorithms follow a consensus-based approach in which 
coordination information is exchanged among controllers to reach 
a common network state. This induces some delay that, as shown
in Sec.~\ref{sec:aut}, can heavily affect the controller reactivity 
perceived at the switches. Indeed, any read/write of a shared data 
structure at a controller is directed to a possibly different 
``data owner'' controller. 
In this case, the controller-to-controller delays must be added 
to the switch-to-controller delays when 
evaluating the controller's reactivity perceived at the switches.   
Thus, the optimal placement of the controllers on the network topology 
must consider not only the delays between the switches and the controllers, 
but also the delays between controllers.
As specified in Sec.~\ref{sec:pw}, most of the past literature concentrated 
on the Openflow-based interaction, thus considered the switch-to-controller delays only, 
and neglected the controller-to-controller delays, which are instead 
the focus of our work.

\ifconf\else
{\color{blue}
The adoption of the Software Defined Networking paradigm 
in Wide Area Networks (SDWANs), poses severe technical challenges. 
Indeed, the design of the network supporting the SDN control plane in WANs 
is more challenging than in data centers, where the limited physical 
distances between network devices permits the installation of a separated 
and dedicated network among the controllers (e.g., \ using Ethernet or 
InfiniBand connections), providing an out-of-band control plane. 
Furthermore, communication delays  are typically negligible in data centers.
Conversely, the problem of supporting a responsive controller-to-controller interaction 
becomes of paramount importance for SDWANs, due to their geographical extension.
Furthermore, SDWANs are based on an in-band control plan: the control packets 
and data packets share the same network infrastructure. 
Separation mechanisms are normally used (e.g., by defining 
one dedicated VLAN for control traffic and/or possibly exploiting priorities)
to run on the same physical infrastructure two logical networks, one for the data 
and one for the SDN control messages. 
}
\fi


In this paper, we provide the following novel contributions:
\begin{enumerate}
\item we evaluate the reaction time perceived at the switches 
when interacting with the controllers, due to the inter-controller 
control traffic, and we prove the relevant role of the adopted 
data-ownership model;
\ifconf
\else
\item {\color{blue}we validate the previous findings with 
traffic measurements on an operational SDWAN;}
\fi
\item we discuss Pareto-optimal controller placements considering 
controller-to-switch and controller-to-controller delays for 
WAN topologies adopted in some real ISP 
\ifconf
\else
networks;
\item {\color{blue}we propose a low-complexity algorithm 
to find the approximated Pareto frontier in large networks.}
\fi
\end{enumerate}



The paper is organized as follows. 
In Sec.~\ref{sec:dc} we provide an overview of distributed SDN architectures. 
We describe the interaction in the control plane, highlighting 
the role of the controller-to-controller communications. 
In Sec.~\ref{sec:aut} we define 
the data-ownership models and the controller placement problem. We propose an analytical model to evaluate the reaction time for the different data-ownership 
\ifconf
\else
{\color{blue}
models and, to show its wide applicability, we apply it to a specific forwarding application in OpenDaylight. 
The model is experimentally validated in Sec.~\ref{sec:validation}. }\fi
In Sec.~\ref{sec:place} we discuss the controller placement problem and present the numerical results obtained by optimizing the controller placement in realistic ISP topologies.
\ifconf
\else {\color{blue}To solve the optimal placement problem in large networks,  in Sec.~\ref{sec:algo} we propose an evolutionary algorithm and investigate its performance.} 
\fi
In Sec.~\ref{sec:pw} we discuss the related work.
Finally, in Sec.~\ref{sec:con} we draw our conclusions.

%% file: controllers.tex

\section{Distributed SDN controllers}\label{sec:dc}
\ifconf 
\else
{\color{blue}Many architectures have been proposed to support distributed SDN controllers, 
to improve the performance and scalability of SDN networks and/or to ensure 
reliability. We concentrate on some specific architectural aspects only. 
For more details, the reader can refer to the bibliography cited in~\cite{survey15}.}
\fi
In distributed controllers, two control planes can be identified. 
First, the switch-to-controller plane, denoted as {\em Sw-Ctr plane}, 
supports the interaction between any switch and its controller (denoted as {\em master controller}) through 
the controller's ``south-bound'' interface. This interaction is usually devoted 
to issue data plane commands (e.g., through the OpenFlow (OF)~\cite{of} protocol) and to configure and 
manage network switches (e.g.\ through OF-CONFIG or OVSDB protocols). 
Second, the controller-to-controller plane, denoted as {\em Ctr-Ctr plane}, 
permits the direct interaction among the controllers through the controller's 
``east-west'' interface. Indeed, the controllers 
exchange heart-beat messages to ensure liveness and to support resilience 
mechanisms. Controllers need also to {\em synchronize the shared data structures} 
to  guarantee a consistent global network view.

Instead, the traffic in the Sw-Ctr plane mainly depends  on the network application 
running on the controller. 
For example, for a reactive application, a {\tt packet-in} message with a copy of the first   
packet of a new flow is sent from the switch to the controller, which replies usually with 
a {\tt flow-mod} to install a flow-specific forwarding rule. After such reply, 
the following packets of the same flow can be directly forwarded 
by the switch to the destined port without interaction with the controller. 
As a consequence, the {\em reactivity of the controller}, defined as the latency perceived by the switch to install the forwarding rule for a new flow, is lower 
bounded by the round trip time between the switch and its master controller. 

\subsection{Data consistency models} \label{sec:raft}

The traffic on the Ctr-Ctr plane is instead crucial to achieve a consistent 
shared view of the network state, which is the required condition to run correctly 
network applications.
The network state is stored in shared data structures (e.g., topology graph, the mapping between any switch to its master controller, the list of installed flow rules), whose consistency across the SDN controllers can be either {\em strong} or {\em eventual}. 
Strong consistency implies that contemporary reads of some data occurring in different controllers always lead to the same result. Eventual consistency implies that contemporary reads may eventually lead to different results, for a transient period. 
Different levels of data consistency 
heavily affect the availability and resilience of the controller, as 
the well-known CAP~theorem highlights~\cite{cap,capnet}.
\ifconf
\else
{\color{blue} 
In a nutshell, anytime a data structure is shared across 
the controllers, they must synchronize through a consensus algorithm 
that guarantees a consistent view of the data in case of updates. 
In general, a consensus algorithm is  very complex, to deal with all possible 
failures and network partitions, and it is tailored to a specific level of 
data consistency. Each controller is required to interact with the other 
controllers through the Ctr-Ctr plane, thus introducing some latency 
to synchronize their internal data structures.}
\fi

In both OpenDaylight (ODL)~\cite{opendaylight} and Open Network Operating System (ONOS)~\cite{onos}, 
two of the most relevant SDN controllers, strong consistency for the shared data structures is achieved by  the recently proposed Raft consensus algorithm~\cite{raft}.
\ifconf\else
{\color{blue}
Indeed, the most recent versions of ODL  (e.g., Beryllium) provide a clustering service to support multiple instances of the 
controller, and the clustering module can be built with a Raft 
implementation~\cite{clustering}, whose code is available in~\cite{raft-code}. 
ODL clustering service organizes the data of different modules into shards. Each shard
is replicated to a configurable odd number of ODL 
controllers. 
Similarly, all the most recent versions of ONOS (2015-16) adopt the Raft algorithm 
for distributed data stores creation  and mastership maintenance~\cite{onos-roadmap}, according to which data is shared across different partitions. }
\fi
%
\ifconf
\else
{\color{blue}

Raft consensus algorithm} 
\fi 
is based on a logically centralized approach, since any data update 
is always forwarded to the controller defined as {\em leader} of the data structure.
Then, the leader propagates the update to all the other controllers, defined as {\em followers}. 
The update is considered committed whenever the majority of the follower controllers 
acknowledges the update. 
\ifconf
\else
{\color{blue}
Sec.~\ref{sec:sd} will describe the adopted protocol, based on the details of the algorithm provided in ~\cite{raft}. }
\fi
Note that the role of master/follower  controller for some data structure is in general independent of the  role of master/slave controller for a switch.

In ONOS data can be also synchronized according to an eventual consistent model, in parallel to strong-consistent data structures. Eventual consistency is achieved through the so called ``anti-entropy'' algorithm~\cite{icc16} according to which updates are local in the master controller and propagate periodically in the background with a simple gossip approach: each controller picks at random another controller, compares the replica and eventual differences are reconciled based on timestamps.

\section{Data-ownership and reactivity in distributed controllers}\label{sec:aut}
 
The controller reactivity as perceived by a switch depends on the local availability 
 of the data necessary for the controller. We can identity two distinct operative models.

In a {\em single data-ownership} (SDO) model,  a single controller (denoted as ``data owner'') 
is responsible for the actual update of the data structure, and any read/write 
operations on the data structures performed by any controller must be forwarded to the data owner.
In this case, the Ctr-Ctr plane plays a crucial role for the  interactions 
occurring in the Sw-Ctr plane, because some Sw-Ctr request messages (e.g., 
{\tt packet-in}) trigger transactions 
with the data owner  on the Ctr-Ctr plane. Thus, the perceived controller reactivity
is also affected by the delay in the Ctr-Ctr plane.
As discussed in Sec.~\ref{sec:raft}, this data-ownership model is currently adopted in ODL and ONOS, for all the strong-consistent data structures managed by Raft algorithm: 
a local copy of the main data 
structures is stored at each controller, but any read/write operation is always 
forwarded to the leader. With this centralized approach, data consistency 
is easily managed and the distributed nature of the data structures is exploited only during failures.

In a {\em multiple data-ownership} (MDO) model, each controller has a local copy 
of the data and can run locally read/write operations. A consensus algorithm 
distributes local updates to all the other controllers. This model has 
the advantage of decoupling the interaction in the Sw-Ctr plane from the one 
occurring in the Ctr-Ctr plane, thus improving the reactivity perceived 
by the switch. The main disadvantage is the introduction of possible 
update conflicts that must be solved with ad-hoc solutions and of possible temporary data state inconsistencies leading to network anomalies (e.g.\ forwarding loops)~\cite{capnet}. Thus, the model applies to generic eventual consistent data structures, as the ones managed by the anti-entropy algorithm in ONOS.

\begin{figure}[!tb]
  \centering
 \includegraphics[scale=0.2]{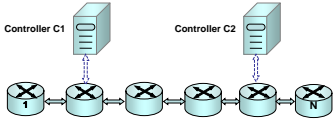}
  \caption{Placement with minimum Sw-Ctr delay}
  \label{fig:toy1}
\end{figure}
\begin{figure}[!tb]
  \centering
 \includegraphics[scale=0.2]{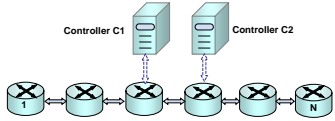}
  \caption{Placement with minimum Ctr-Ctr delay}
  \label{fig:toy2}
\end{figure}

We concentrate our investigation on the delay tradeoff achievable in the Sw-Ctr and in the Ctr-Ctr control planes. 
For the MDO model, the two planes are decoupled, as shown later in Property~\ref{prop:1d}. Thus, small Sw-Ctr delays imply high reactivity of the  controllers (i.e.\ small reaction time), whereas small Ctr-Ctr delays imply lower probability of network state inconsistency.
For  the SDO model,  Property~\ref{prop:1s} will show that the Ctr-Ctr delays affect not only the resilience but also the perceived reactivity of the controllers. Thus, reducing Ctr-Ctr delays is important as reducing Sw-Ctr delays; but, for topological reasons, reducing one kind of delays implies maximizing the other, and vice versa.

%
%
Indeed, consider the toy scenario depicted in Figs.~\ref{fig:toy1}-\ref{fig:toy2}, comprising  $N$ switches in a linear topology. We assume that each switch selects the closest controller as its master and that the delays between two nodes are directly proportional to their distance in terms of number of hops. We consider two specific controller placements. In Fig.~\ref{fig:toy1}, the two controllers are placed to minimize the average Sw-Ctr delay, which is (proportional to) $N/8$. The corresponding Ctr-Ctr delay is  $N/2$. 
Instead, in Fig.~\ref{fig:toy2}, the controllers are placed to minimize the Ctr-Ctr delay, which is 1, whereas the Sw-Ctr delay doubles and becomes $N/4$.

We now derive the reactivity for the two data-ownership models.
For simplicity, we consider only the propagation delays of the physical links, and neglect all the processing times and the queueing delays due to network congestion. 

\subsection{Reactivity model for MDO model}

In a MDO scenario, a generic event occurring at the switch (e.g.\ a miss in the flow table) generates a message (e.g., a {\tt packet-in}) to its master controller, which processes the message locally and eventually sends back a control message to the switch (e.g., {\tt flow-mod} or {\tt packet-out} message). In the meanwhile, in an asynchronous way, the master controller advertises the update to all the other controllers. Thus, the reaction time of the controller perceived by the switch, defined as  $T_R^{(m)}$, can be  evaluated as follows:
\begin{prop}\label{prop:1d}
In a MDO scenario for distributed SDN controllers,  
the reaction time perceived at the switch is: 
 \begin{equation}\label{eq:m}
 T_R^{(m)}=
 2d_\text{sw-ctr}
 \end{equation}
  \end{prop}
  being $d_{\text{sw-ctr}}$ the delay from the switch to its master controller.

\subsection{Reactivity model for SDO model}\label{sec:sd}

\begin{figure}[b!]
\centering
\includegraphics[width=1\linewidth]{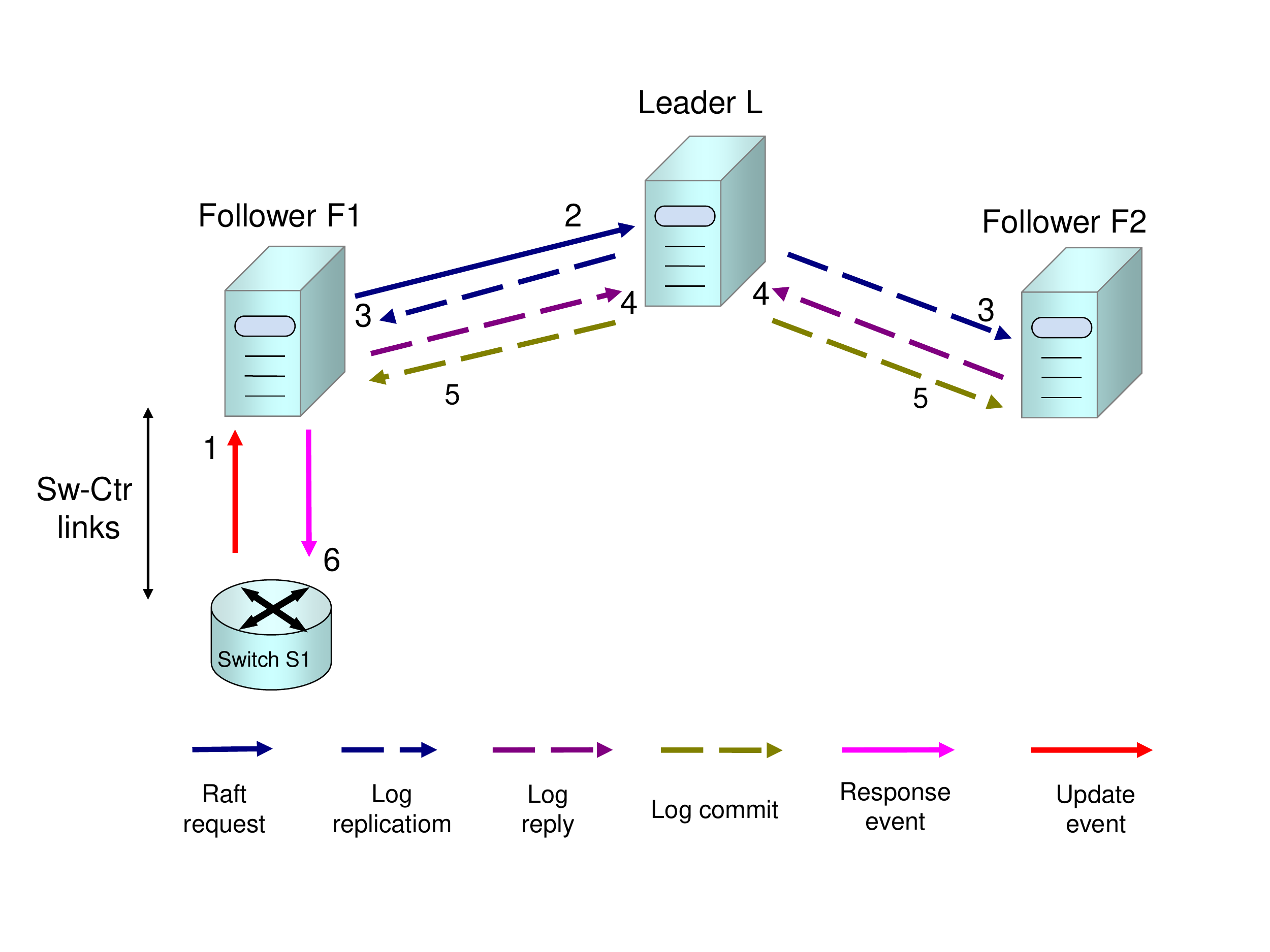}
\caption{Control traffic due to SDO model for an update event at the switch.\label{fig:ODL1}}
\end{figure}

%
%

In a SDO scenario, we assume the exchange of messages coherent with the detailed description of Raft algorithm available in~\cite{raft} and devise a model to evaluate the reactivity of the controller as perceived by the switch.  
\ifconf
\else
{\color{blue}
This analytical model, here obtained in a speculative way, will be later tailored to a specific ODL network application in Sec.~\ref{sec:model}
and experimentally validated in a real SDN WAN networking running ODL (see Sec.~\ref{sec:validation}).}
\fi

\ifconf
\else
{\color{blue}
According to Raft implementation in ODL, the controller can operate as either unique leader or as one of the followers, for a specific data store (denoted ``shard'' in the following).  
As an example, Fig.~\ref{fig:ODL1} shows a general message exchange sequence in a cluster with 3 controllers (one leader and two followers), when an update event (e.g.\ {\tt packet-in} message) for the shard is generated at some switch (S1 in the figure), which receives a response message from its controller due to the update (e.g.\ {\em packet-out} message). 
The arrows show the exchange of messages in both Sw-Ctr and Ctr-Ctr planes triggered by the update event; the number associated to each arrow shows the temporal sequence of each packet. 
We have now two cases. 

In the first case,  S1's master controller is a follower for the shard (as depicted in Fig.~\ref{fig:ODL1}). Thus,  
the switch sends the update event (message 1) to the master controller, which asks the leader to update the shard through a ``Raft request'' (message 2). Now the leader sends a ``log replication'' message to all its followers (message 3) and waits for the acknowledge from the majority of them (``log reply'' in  messages 4). Only at this point, the update is committed through a ``log commit'' (message 5) sent to all the followers. Thus, after receiving the commit message, S1's master controller can process the  update on the shard and generate the response event (message 6) to the switch.  

In the second case, S1's master controller is the leader for the shard. This case is identical to the previous one except for the ``Raft request'' message 2, now missing.}

\fi
\ifconf
\else
{\color{blue}For both cases, the}
\fi 
{\em controller reaction time} perceived by switch S1 is given by the time between the update event and the response event messages.
%
 Let $d_\text{sw-ctr}$ be the communication delay between the switch and its master controller and $d_\text{ctr-leader}$ the communication delay from the master controller and the leader (being null whenever the master is also leader). Assume a cluster of $C$ controllers. Because of the majority-based selection, let $d_\text{ctr*-leader}$ be the communication delay between the leader and the farthest follower belonging to the majority (i.e.\ corresponding to the $\lfloor(C/2)+1 \rfloor$-th closest follower). 
\ifconf
\else {\color{blue}
Observing Fig.~\ref{fig:ODL1},} 
\fi
the reaction time is obtained by summing twice $d_\text{sw-ctr}$, twice $d_\text{ctr-leader}$ 
\ifconf
\else
{\color{blue}
(only in the first of the above cases)}
\fi and twice $d_\text{ctr*-leader}$. Thus, we can claim:
\begin{prop}\label{prop:1s}
In a SDO scenario (e.g.\ adopting Raft consensus algorithm) for distributed SDN controllers, 
the reaction time $T_{R}^{(s)}$ perceived at the switch is: 
 \begin{equation}\label{eq:1s}
 T_{R}^{(s)}=
 2d_\text{sw-ctr}+2d_\text{ctr-leader} + 2d_\text{ctr*-leader}
\end{equation}
\end{prop}
\ifconf
\else {\color{blue}
Thus, the reaction time is identical to the one for MDO model plus either 2 or 4 times the RTT between the controllers, when the master controller is either leader or follower of the shard, respectively.
Notably, the delays between controllers may be dominant for large networks as SDWAN, as shown experimentally in Sec.~\ref{sec:validation}.}
\fi

\ifconf
\else
{\color{blue}
\subsection{Reaction time for reactive forwarding in ODL}\label{sec:model}

To show the wide applicability of the SDO model devised in Sec.~\ref{sec:sd} and its practical relevance, we apply Property~\ref{prop:1s} to compute the flow setup time for the specific 
layer-2 forwarding application called ``l2-switch'' available in ODL, deployed on a generic topology.
Notably, even if this analytical model is derived in a speculative way, in Sec.~\ref{sec:validation} we will show that it is {\em very accurate} from experimental point of view, and thus its relevance is practical. The same methodology can be applied to  analyze other applications, given the knowledge of their detailed behavior.

 L2-switch application provides the default reactive 
forwarding capabilities and mimics the learning/forwarding mechanism at layer 2 of standard Ethernet switches.
Anytime a new flow enters the first switch of the network, the corresponding ARP-request is flooded to the destination, and only when the ARP-reply is generated, the controller installs a forwarding rule at  MAC layer in all the switches involved in the path from the source to the destination, and vice versa. The  association of a MAC address to the switch port, needed for the learning phase, 
is distributed to the other controllers using Raft algorithm. 

\begin{figure}[tb!]
\centering
\includegraphics[width=1\linewidth]{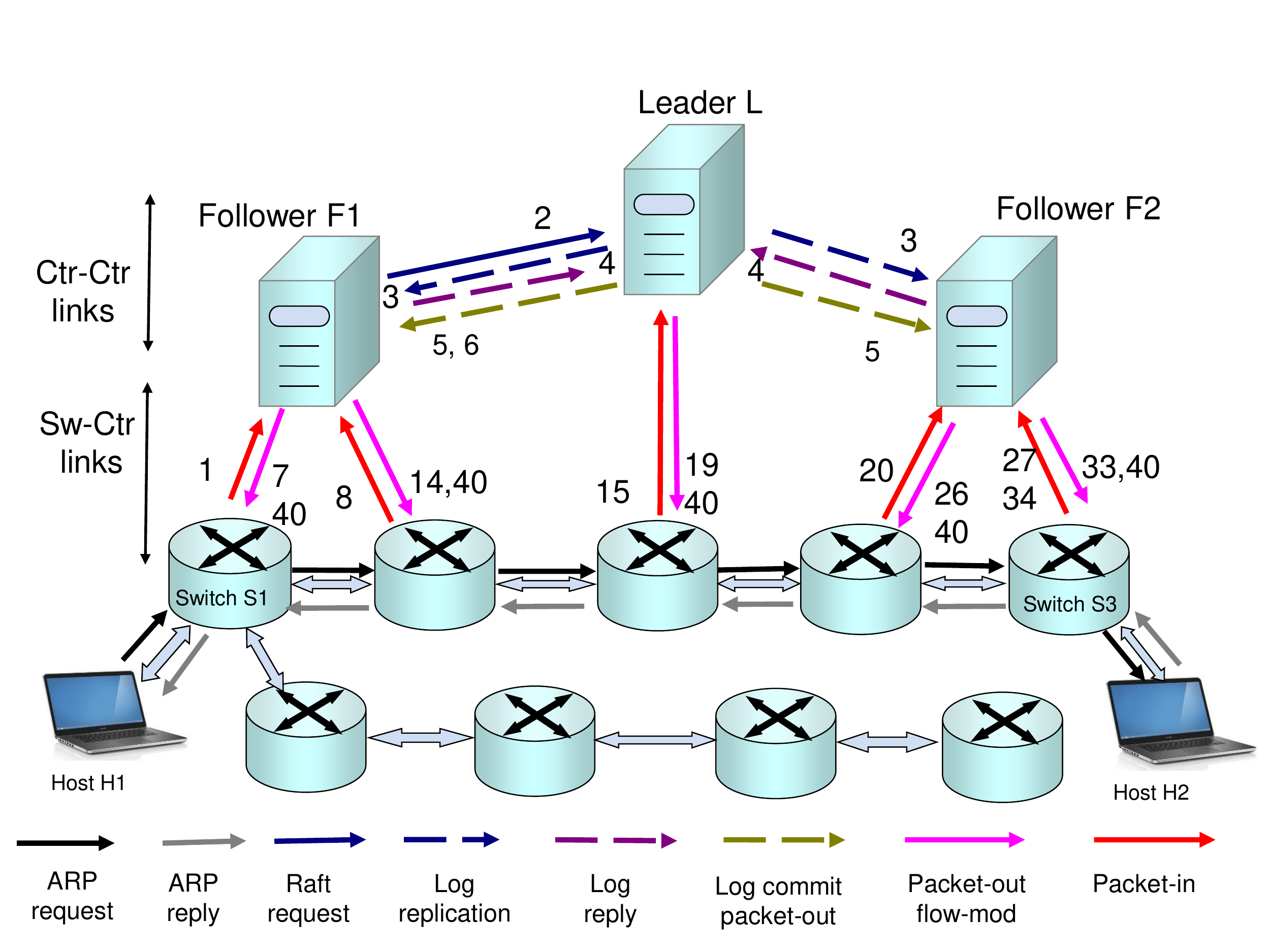}
\caption{The control traffic for l2-switch application in ODL. For the sake of clarity, we report just some sequence numbers. The packets sent with sequence 2-6 repeat as 9-13, 21-25, 28-32, 35-39. The packets sent with sequence 3-5 repeat as 16-18, since the master controller of the third switch along the path is also the shard leader.
Only the messages between the controllers and the switches along the source-destination path are shown. }
\label{fig:ODL2}
\end{figure}

Assume a generic topology as shown in Fig.~\ref{fig:ODL2} connecting source host H1 to destination host H2, with every switch $s$ attached to its master controller (denoted as $c(s)$) which can be either a follower or a leader (denoted as $L$) within the cluster.
We assume initially empty flow tables in all the switches. 
At the beginning, the first ARP-request corresponding to a new flow from H1 is flooded in the whole network (avoiding loops by precomputing a spanning tree on the topology). Anytime the ARP packet is received at a switch, a packet-in is generated and the association (MAC source address, ingress port, switch identifier) is stored in the shared data store, in order to mimic the standard learning process. This means that at each switch, along the path from the source to the destination, a latency is experienced according to \eqref{eq:1s} of Property~\ref{prop:1s}. When the ARP-request reaches the destination, H2 sends back an ARP-reply which generates a packet-in from the last switch (denoted as $s'$) to the controller. This event generates another update since the controller learns the port of $s'$ at which H2 is connected. Only at this point, the controller installs a flow rule across all the switches in the source-destination path and then the ARP-reply is switched back to the source. Thus the flow setup time can be evaluated as {\em ARP reaction time} $t_{\text{r}, ARP}$, defined as the time between the sent ARP-request and the received ARP-reply, both evaluated at the source host.
Let $d_{i,j}$ be the propagation delay between nodes $i$ and $j$. Let $\mathcal P$ be the list of all the nodes involved in routing path from $H1$ to $H2$, in which the last switch  $s'$ appears twice. Thus, the total number of  updates within a flow is $|\mathcal P|$.
We can claim:
\begin{prop}\label{prop:odl}
In OpenDaylight (ODL) running l2-switch application, the flow setup time (ARP reaction time) can be computed as
\begin{multline} \label{eq1}
t_{\text{r},ARP} = 
2d_{H1,H2}+\sum_{s\in\mathcal P} (2d_{s,c(s)}+2d_{c(s),L})+\\ 
2|\mathcal P| d_{\text{cnt*-follower}}+
|\mathcal P| t_\text{c}
\end{multline}
\end{prop}
Indeed, the first term in~\eqref{eq1} represents the delay  to send the ARP request and reply along the routing path, the second term represents the delay occurring for all the switches along the path (the final switch $s'$ is double counted) due to the packet-in and the packet-out/flow-mod ($2d_{s,c(s)}$) and due to the Raft-request and log commit ($2d_{c(s)L}$), the third term represents the delay to get the acknowledgement from the majority for each of the updates, and the fourth term represents the computation time for each update at the controller (assuming to be constant and equal to $t_\text{c}$).
}
\fi

%% file: validation.tex
\ifconf
\else
{\color{blue}
\section{Experimental validation of the single data-ownership (SDO) model}\label{sec:validation}


 We validated the analytical model proposed in Sec.~\ref{sec:sd} on a real and operational network.  Specifically, we run a cluster of OpenDaylight (Helium SR3 release) controllers running the default Simple Forwarding Application.  
 We run our experiments in the JOLNet, which is an experimental SDN network deployed  by Telecom Italia Mobile (the major telecom operator in Italy). JOLnet  is an Openflow-based SDWAN, with 7 nodes spread across the whole Italy, covering Turin, Milan, Trento, Venice, Pisa and Catania. Each node is equipped with an OF switch and a compute node. 
The compute node is a server deploying virtual machines (VMs), orchestrated by OpenStack. Network virtualization is achieved though FlowVisor~\cite{fv} and  the logical topology among the OF switches is fully connected.

\subsection{Methodology for the validation}

Due to the limited number of nodes in the JOLNet and the limited flexibility in terms of topology, we augmented the topology with  an emulated  network running on Mininet~\cite{mininet} in one available compute node. 
We adopted the linear network topology of Fig.~\ref{fig:linear} with a variable number of nodes (from 3 to 36) and with one host attached at each switch. We generated ICMP traffic using the ping command among the different hosts. 
We run a single controller cluster with multiple ODL instances running in different nodes of the JOLNet,  in order to distribute geographically the controllers across Italy. 
The controllers instances and Mininet run individually on single VMs for a flexible placement across the nodes. 
Thanks to the large physical extension of the network, we could experiment a large variety of  scenarios, e.g.\ with large Sw-Ctr delays (when the VM of the master controller was located in a far compute node from the switch node) and/or large Ctr-Ctr delays (when the VMs of the controller instances were located in nodes far one from each other).  
%
By selecting the master controller of the switches, we were able to change the data owner of the shared data structure within the cluster. 
We evaluated the flow setup time in ODL of the Simple Forwarding Application in a cluster of 3 controllers. 


We measured the flow setup time $t_\text{r}$ as defined in Sec.~\ref{sec:model}. As a reminder, it 
is the delay  perceived by a flow at the first switch, namely the latency between the time when the first switch sends the packet-in message to controllers and the time it receives back the packet-out/flow-mod messages.
This latency was measured by comparing the timestamps of the packets obtained by using Wireshark as network sniffer at the Mininet interface towards the controllers.


\begin{figure}[!tb]
	\centering
	\includegraphics[page=1, width=0.9\columnwidth]{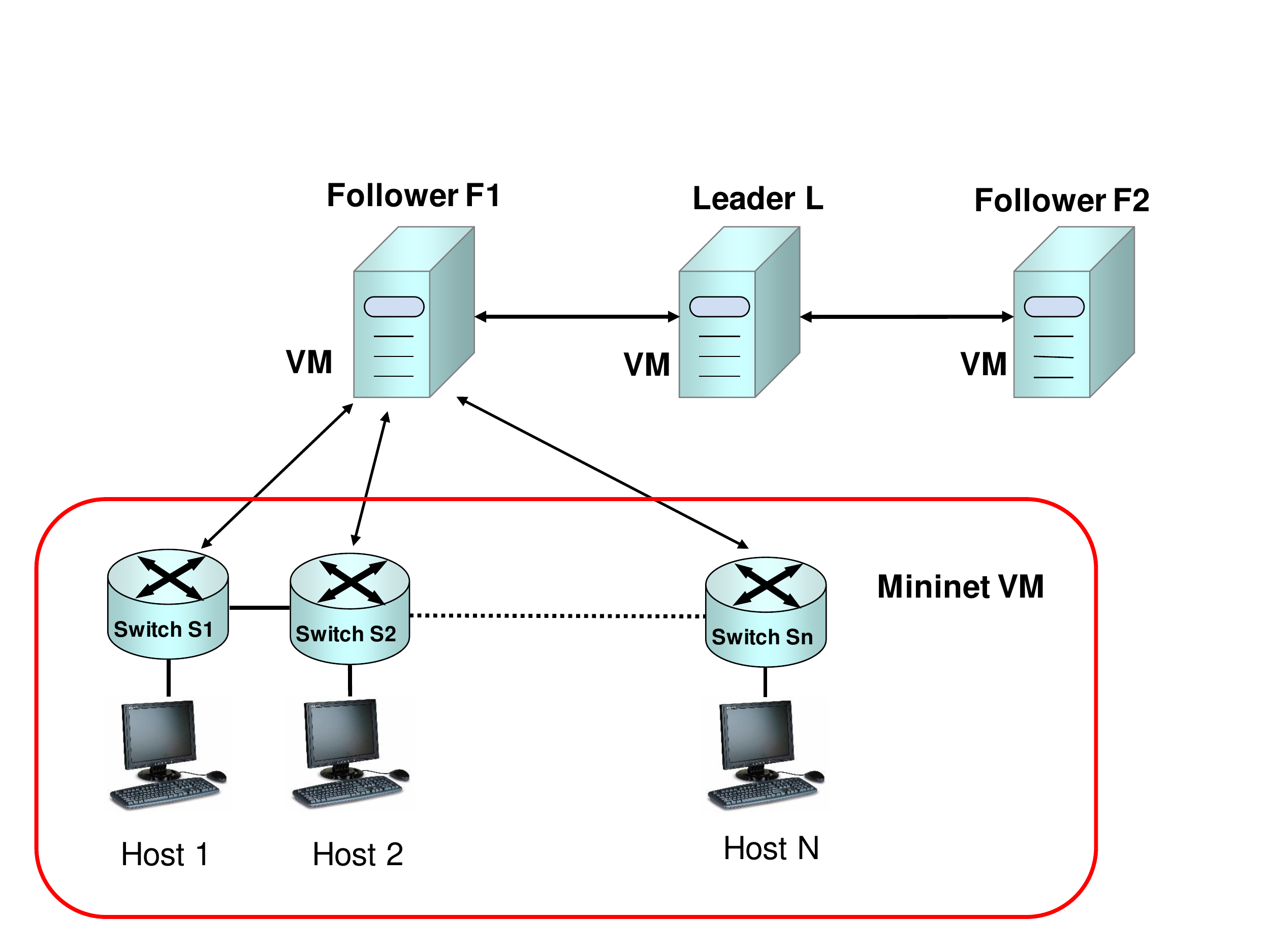}
	\caption{Network configuration for the validation of the SDO model}
	\label{fig:linear}
\end{figure}

As first step to validate Property~\ref{prop:odl}, we evaluated the RTT among each pair of nodes in JOLNet and thus we were able to compute the delay between any pair of nodes $i$ and $j$ as $d_{ij}=\text{RTT}_{ij}/2$, required to apply~\eqref{eq1}. The experiments reported in the following refer to the scenarios using 3 JOLnet nodes, namely Turin, Milan and Pisa, to deploy the VMs. 
The RTT between these nodes are shown in Fig.~\ref{fig:rtt}.

As second step, in order to effectively validate Property~\ref{prop:odl}, we performed multiple measurements (typically, 100) by clearing the whole forwarding tables and restarting the controllers at each run. We evaluated the 95\% confidence intervals of the measurements.
The {\em accuracy of the measurements} were computed according to the formula:
\(
\lambda = {I_{95}}/{(2\bar{\mu})}
\)
where $I_{95}$ represents the width of the 95\% confidence interval and $\bar{\mu}$ the average measure. 
For each scenario and topology, the {\em relative error of the model} was computed as: 
\(
\delta = {|M_i - T_i|}/{|T_{i}|}
\)
where $M_i$ is the average flow setup time according to the experiments and $T_i$ is the flow setup time according to  Property~\ref{prop:odl}.

We considered different scenarios, depending on the placement of the controllers and of Mininet across the different JOLNet nodes. We  refer to the physical distance between the network nodes (emulated with Mininet) and the controllers (followers and leader) as ``close'' when the corresponding VMs are running in the same node, and ``far'' on remote nodes. Table~\ref{table-scenaria} lists all the experimented scenarios, discussed in the following section. In our cluster of 3  ODL controllers, the leader controller is denoted as ``L'' and the two followers are denoted as ``F1'' and ``F2''. Controller F1 is set to be master controller for all the switches in  Mininet network. ``Net'' represents Mininet emulated network.

\subsection{Experimental results}

\begin{table}
	\renewcommand{\arraystretch}{1.3}
	\caption{Placement of the VMs for the experimented scenarios. ``L'': leader controller. ``F1'', ``F2'': follower controllers. ``Net'': Mininet.}
	\label{table-scenaria}
	\centering
	\begin{tabular}{ | c || c | c | c |}
		\hline
		Scenario & Turin & Milan & Pisa \\ 
		               & VMs & VMs & VMs \\ 
		\hline 
		TT & Net, L, F1, F2 & - & - \\
		TMC & Net, F1 & L, F2 & - \\ 
		TMF & Net & L, F1, F2 & - \\ 
		TPC & Net, F1 & - &  L, F2 \\ 
		TPF & Net & - & L, F1, F2 \\
		\hline
	\end{tabular}
\end{table}

\begin{table}
\caption{Input parameters for the model, accuracy of measurements and relative error of the model}\label{tab:res}
	\centering
	\begin{tabular}{ | c ||c|c|c||c|c|}
	\hline
	Scenario & $d_\text{sw-ctr}$ & $d_\text{ctr-ctr}$ & $t_c$ & Experimental & Model  \\
	                &                &                             &                             &   accuracy ($\lambda$) & error ($\delta$) \\
	\hline
	TT & 0.25~ms& 0.25~ms & 20~ms & 1.2\% - 2.7\% & 3.2\%\\
	TMC & 0.25~ms & 2.0~ms & 20~ms & 0.7\% - 3.9\%& 5.2\% \\
	TMF & 2.0~ms & 0.25~ms & 20~ms &  0.6\% - 3.6\% & 5.1\% \\ 
	TPC & 0.25~ms & 66~ms & 20~ms& 0.3\% - 1.3\% & 9.2\% \\
	TPF & 66~ms & 0.25~ms & 20~ms & 0.6\% - 2.3\% & 0.5\% \\
	\hline
	\end{tabular}
\end{table}

\begin{figure}
	\centering
	\includegraphics[page=1, width=4cm]{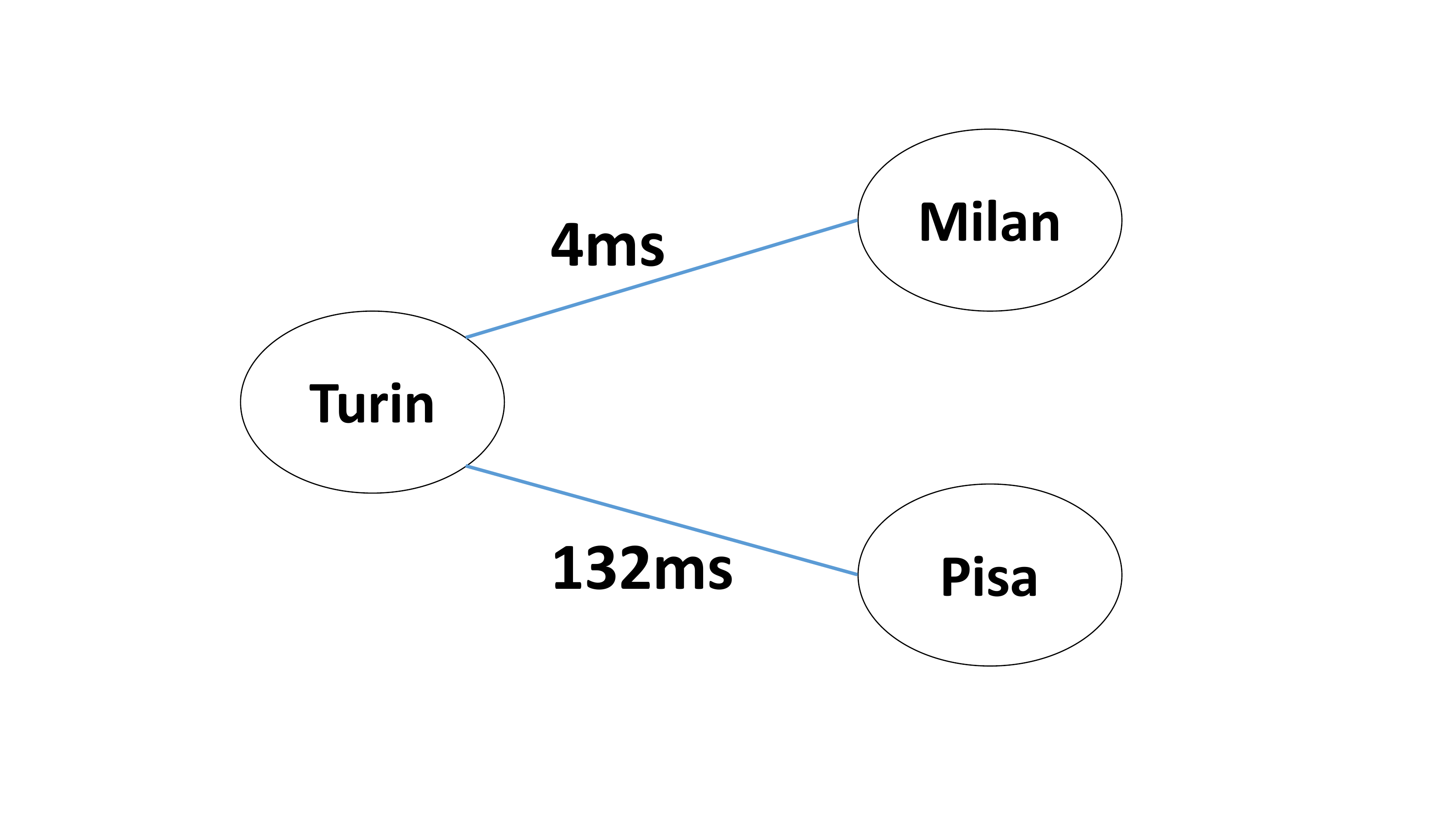}
	\caption{ The average RTT between the JOLNet nodes relevant for the experiments}
	\label{fig:rtt}
\end{figure}

\begin{figure}
	\centering
	\includegraphics[page=1, width=1.010\columnwidth]{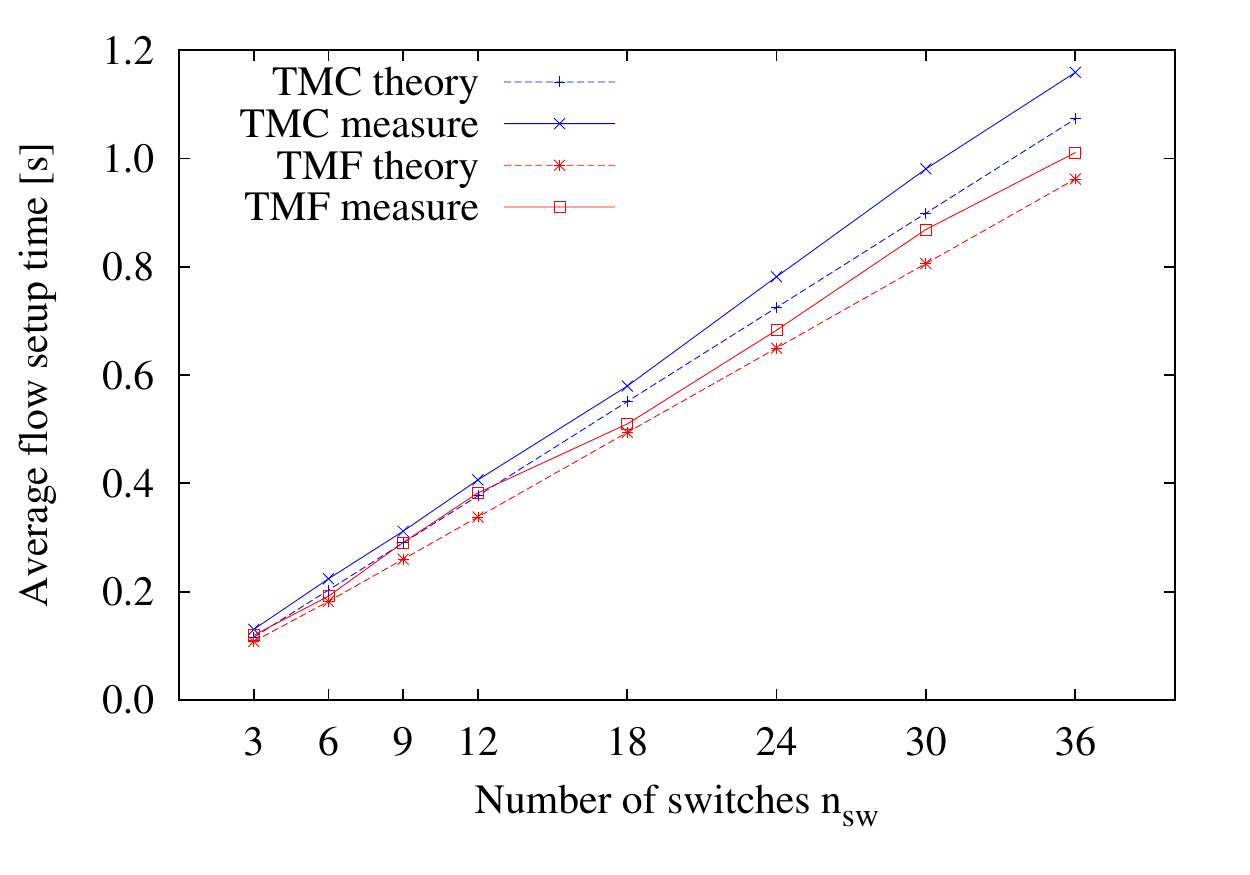}
	\caption{ Experimental results with the VMs running either in Turin or Milan }
	\label{fig:valid1}
\end{figure}

\begin{figure}
	\centering
	\includegraphics[page=1, width=1.010\columnwidth]{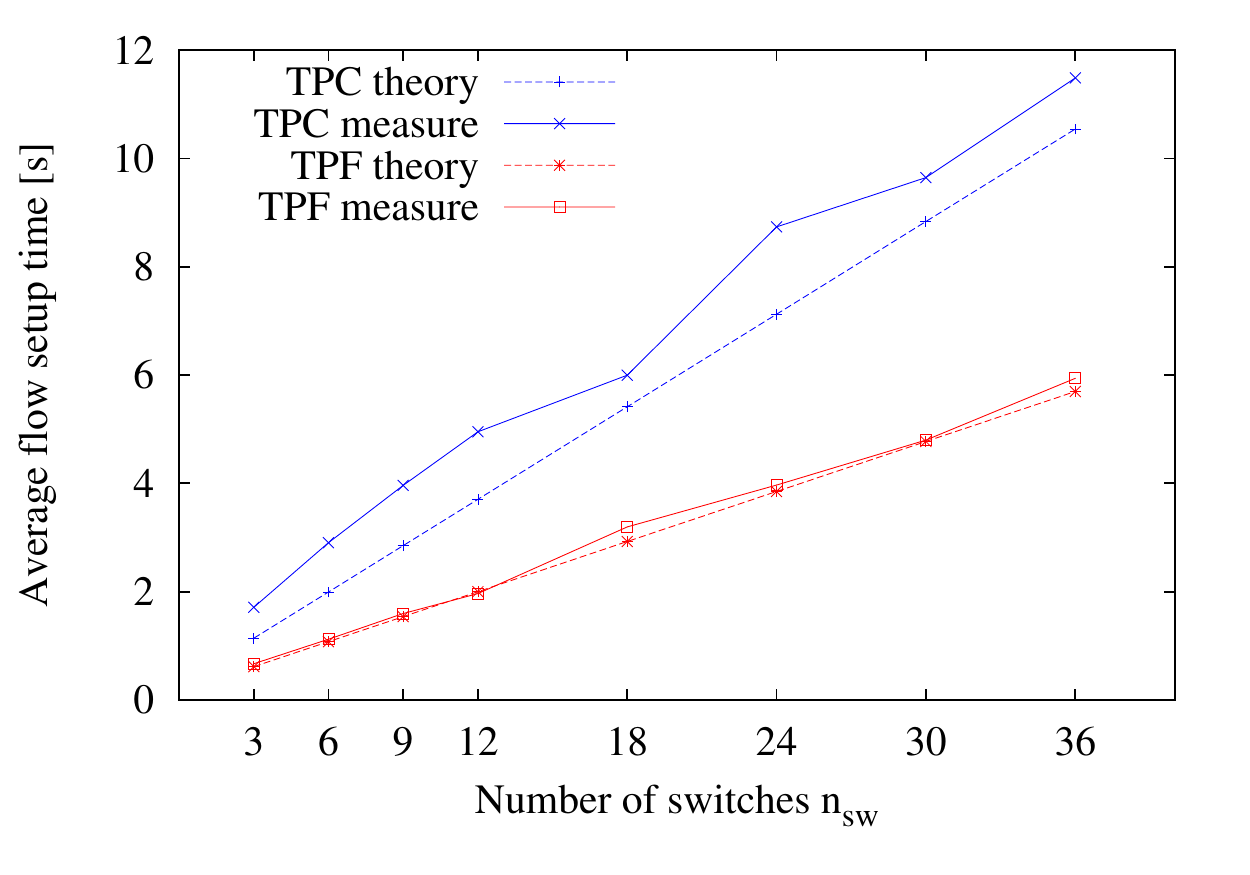}
	\caption{ Experimental results with the VMs running either in Turin or Pisa}
	\label{fig:valid2}
\end{figure}



Table~\ref{tab:res} summarizes the input parameters  that  have been used in~\eqref{eq1}, and shows also the final experimental results in terms of the accuracy of the measured values and the model error. The input parameters have been obtained either by the measurements in Fig.~\ref{fig:rtt} when the VMs were located in different nodes, or by following the steps explained below. In more detail:
\begin{itemize}[\setlength{\IEEElabelindent}{\dimexpr-\labelwidth-\labelsep}
    \setlength{\itemindent}{\dimexpr\labelwidth+\labelsep}
    \setlength{\listparindent}{\parindent}]
\item {\bf Scenario TT (Turin-Turin):}
We run the VMs of all the controllers and of Mininet in the same node, in order to evaluate the baseline latency due to the controller processing time and to the communication overhead (through the local virtual interfaces).  
First, we measured the communication delay between VMs, due to the local hypervisor running the different VMs, using ping command. We measured 0.5~ms, thus the delay between the network and the controller, and between controllers was assumed to be 0.25~ms.
By running Mininet and measuring the flow setup time, we estimated an average processing latency of 20~ms, used as reference for all the other experiments.
We run many experiments varying $n_\text{sw}$ in the interval $[3,36]$ and observed an average relative error 3.2\% of the model expressed by~\eqref{eq1} with respect to experimental data. 


\item {\bf Scenario TMC (Turin-Milan-Close):}
In this scenario, leader L and Follower F2  of the cluster are located in Milan, whereas follower F1 is co-located with the network in Turin node, thus all OF switches are close to their master controllers. 
The dominant term in~\eqref{eq1} is the delay between controllers, equal to $4/2=2$~ms.
Fig.~\ref{fig:valid1} shows the average flow setup time computed according to~\eqref{eq1} and the one measured. According to Table~\ref{table-scenaria}, the experimental results were quite stable for the different  values of nodes. 
The measured value was quite close to the theoretical one, with a relative error 5.2\%.
Interestingly, the flow setup time in absolute values is always larger than 100~ms and can reach also 1.2~s when the network is quite large. Notably, this is mainly due to the interaction between the master controller and the leader controller.  
\item {\bf Scenario TMF (Turin-Milan-Far):}
All the controllers are located in Milan, thus all OF switches are far from their master controller. Thus, now the dominant term in~\eqref{eq1} is  the delay from the switch to the controller (2~ms). Fig.~\ref{fig:valid1} compares the theoretical flow setup time with the measured ones. Now the relative error of the model is 5.1\%.
As in the previous case, the flow setup time can be very large, due to the latency in the interaction between the network and its master controller.

\item {\bf Scenario TPC (Turin-Pisa-Close)} 
All OF switches and their master controller F1 are located in Turin, whereas the leader controller and the other controller are located in Pisa. The dominant term is the delay between controllers ($132/2 = 66 $ ms). As shown in Fig.~\ref{fig:valid2}, the measured value approaches the theoretical value with a relative error of 9.2\%. The measured delays can range from  $2$ to $12$~s as we vary the number of switches. These surprising huge delays are due to the interaction between leader L and follower F1.

\item {\bf Scenario TPF (Turin-Pisa-Far)}
All the controllers are located in Pisa with the network still in Turin. Due to the large delay between Turin and Pisa (66~ms), the dominant term in~\eqref{eq1} is the delay between follower F1 and the network. In all the results shown in Fig.~\ref{fig:valid2}, the relative error ($0.5\%$) of the model with respect to the theoretical value is very small.  Also in this case, the flow setup time is huge in absolute terms (up to $6$~s),  due to the large delay between the network and the controllers.
\end{itemize}

In summary, our experimental results show clearly that the delays introduced by inter-controller communications can be very large and the great accuracy of our model with respect to experimental data shows exactly the reason for it (i.e.\ the interaction between controllers due to the RAFT consensus algorithm).
}
\fi

%% file: place.tex
\section{The controller placement problem}\label{sec:place}

The Sw-Ctr delays (between the switches and their master controller) and Ctr-Ctr delays (between controllers) have a direct impact on the reactivity of the controller perceived at switch level, as highlighted in Properties~\ref{prop:1d}-\ref{prop:1s}. This observation is particularly relevant for large networks, where propagation delays are not negligible. 
Thus the placement of the controllers in the network is of paramount importance and implies different tradeoffs between Sw-Ctr delays and Ctr-Ctr delays.

Let $N$ be the total number of switches in the network and $C$ be the total number of controllers to place in the topology.
 The output of any placement algorithm can be represented by the vector denoted as {\em placement configuration}:
\begin{equation}
\label{equation1}
\pi=[\pi_c]_{c=1}^C
\end{equation}
where $\pi_c\in\{1,\ldots,N\}$ identifies the switch at which controller $c$ is connected to. We assume that all the controllers are connected to distinct switches (equivalently, two controllers cannot be connected to the same switch), i.e.\ $\pi_{c}\neq \pi_{c'}$ for any $c\neq c'$. 
Let  $\Omega\subset \{1, 2, .., N\}^{C}$ be the set of all placement configurations; thus, the total number of possible placements is
\begin{equation}\label{eq:exa}
|\Omega|={N \choose C}
\end{equation}

The optimal controller placement problem consists of finding $\pi\in\Omega$ such that
some cost function (e.g.\ the maximum or average Sw-Ctr delay) is minimized and it is in general a NP-hard problem for a generic graph, as discussed in~\cite{nick12}.

%% file: methodology.tex
\subsection{Results on the placement of controllers in ISP networks}\label{sec:simu}


To explore all the possible tradeoffs on the  Sw-Ctr  and Ctr-Ctr  planes, we adopt an optimal  algorithm (denoted \textsc{Exa-Place}) to enumerate exhaustively all possible controller  
placements and get all {\em Pareto-optimal} placements\footnote{When considering two performance metrics $x$ and $y$ to minimize, a solution $(x_p,y_p) $ is Pareto optimal if does not exist any other configuration $(x',y')$ dominating it, i.e.\  better in terms of both metrics; thus, it cannot be that $x'\leq x_p$ and $y'\leq y_p$. The set of all Pareto-optimal solutions denotes the Pareto-optimal frontier.} and thus the corresponding Pareto-optimal frontier.
For small/moderate values of network nodes $N$ and number of controllers $C$, as considered in this section, the number of possible placements, evaluated in~\eqref{eq:exa}, is not so large and thus \textsc{Exa-Place}  is computationally feasible.  
\ifconf
\else
{\color{blue}
In Sec.~\ref{sec:algo} we will instead devise an approximated algorithm to find the Pareto frontier for large networks and/or large number of controllers.}
\fi

The network topology is described by a weighted graph  where each node represents a switch; each edge represents the physical connection between the corresponding switches and  is associated with a latency value. Each controller is connected directly to a switch. We assume that the master controller of a switch is the one with the minimum Sw-Ctr delay. We also assume that all the communications are routed along the shortest path.

Coherently with previous work~\cite{nick12}, we have considered specifically the topology available in the {\em Internet topology zoo} website~\cite{zooweb}. 
This repository collects around 250 network topologies of ISPs, at POP level.  For each ISP, the repository provides the network graph, with each node (i.e.\ switch) labeled with its geographical coordinates. From these, we computed the propagation delay between the nodes and associated it as latency of the corresponding edge.
For any given controller 
placement,  we evaluate both the Sw-Ctr delay (as the average delay between the switches and their master controllers) and the {Ctr-Ctr delay} (as the average delay among controllers).

%% file: results.tex
\subsection{Tradeoff between Sw-Ctr and Ctr-Ctr delay}
%


We report only the analysis of three different ISP:  (1) HighWinds, a world-wide network with 18 nodes, 
(2) Abilene, a USA-wide network with 11 nodes, 
(3) York, a UK-wide network with 23 nodes. 
Very similar results have been obtained for other topologies. 

\begin{figure}[!tb]
\centering
\includegraphics[width=8.2cm]{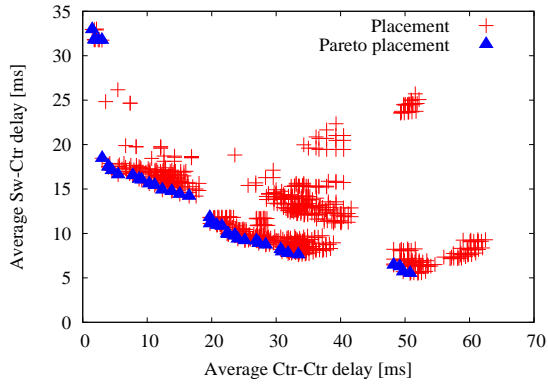}
\caption{Delay tradeoffs  in HighWinds network} 
\label{fig:delay-highwinds}
\end{figure}

\begin{figure}[!tb]
\centering
\includegraphics[width=8.2cm]{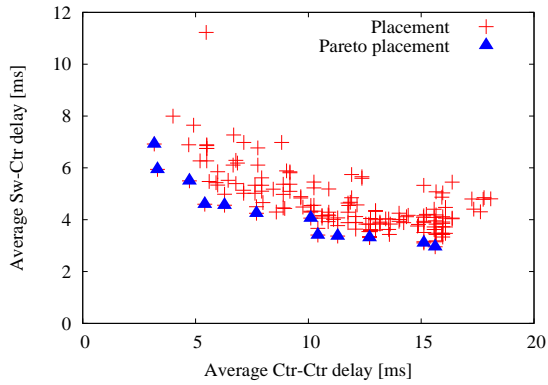}
\caption{Delay tradeoffs in Abilene network} 
\label{fig:delay-abi}
\end{figure}

\begin{figure}[!tb]
\centering
\includegraphics[width=8.2cm]{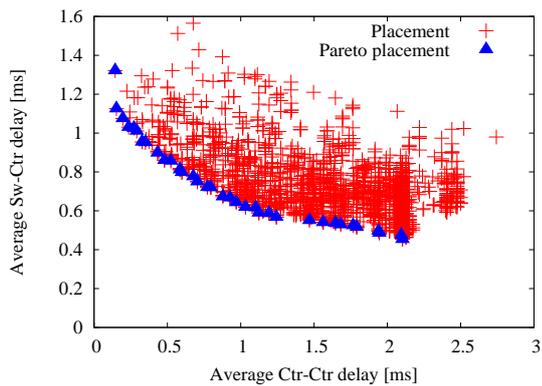}
\caption{Delay tradeoffs in York network}
\label{fig:delay-york}
\end{figure}

 \begin{figure}
 \centering
\includegraphics[page=1, width=8.2cm]{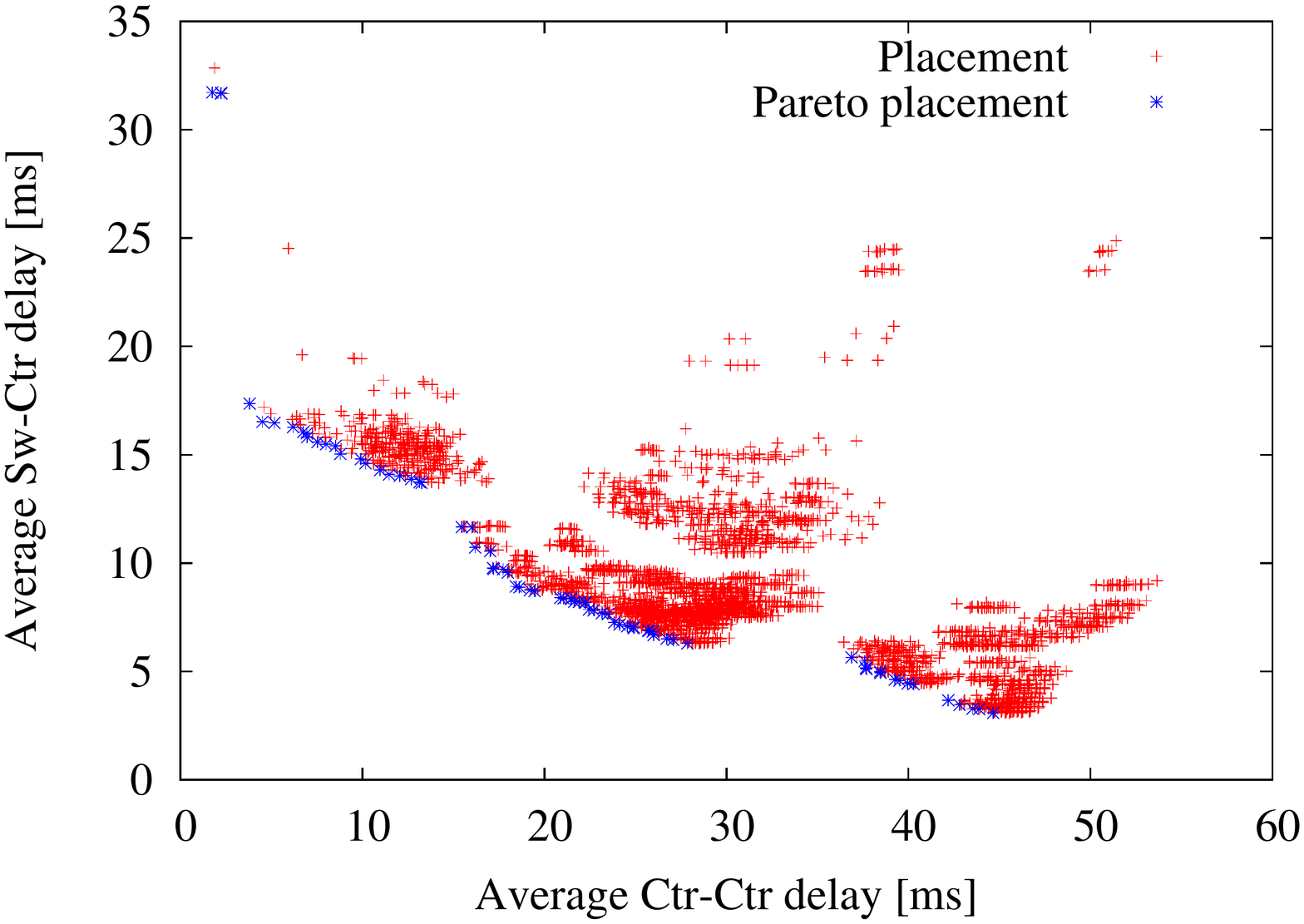}
 \caption{Delay tradeoffs in HighWinds network with 4 controllers}
  \label{fig:delay-highwinds4}
 \end{figure}
 
 \begin{figure}
 \centering
 \includegraphics[page=1,width=8.2cm]{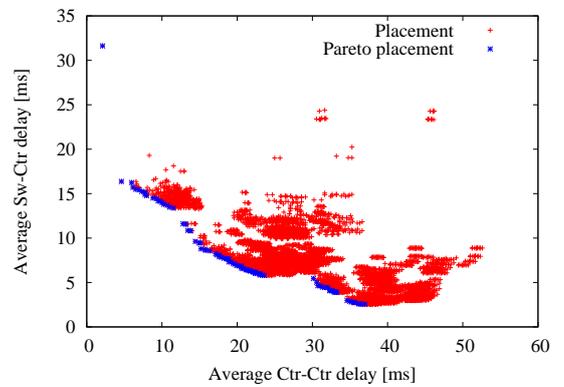}
 \caption{Delay tradeoffs in HighWinds network with 5 controllers}
 \label{fig:delay-highwinds5}
 \end{figure}

Figs.~\ref{fig:delay-highwinds}-\ref{fig:delay-york} show the scatter plot with the Sw-Ctr and Ctr-Ctr delays achievable by all possible placements of 3 controllers, for the three ISPs, respectively. In total, all the possible ${18 \choose 3} = 816$, ${11 \choose 3}=165$ and ${23 \choose 3}=1771$ different placements are shown; the corresponding Pareto-optimal placements are also highlighted.
As observed when discussing the toy example of Figs.~\ref{fig:toy1}-\ref{fig:toy2}, high (or small) Sw-Ctr delays imply small (or high) Ctr-Ctr delays, respectively.
 The graphs show the large variety of Pareto-optimal placements. 
 We denote by $P1$ the Pareto point with the minimum Sw-Ctr delay (i.e.\ the most right-low point), and by $P2$ the one with the minimum Ctr-Ctr delay (i.e.\ the most left-high point).  Table~\ref{tab:gains} shows the delay reduction when we compare $P1$ with $P2$ and can be read as follows:  if we allow the Sw-Ctr delay to increase by the factor shown in the second column, then the Ctr-Ctr delay decreases by the factor shown in the third column. 
Notably, in HighWinds if we allow the Sw-Ctr delay to increase  by 6.0 times, then the Ctr-Ctr delay decreases by  34.8 times, which is very high gain. Also in York the gain is relevant, since an increase in the Sw-Ctr delay by 2.9 times corresponds to a Ctr-Ctr delay reduction of 15.0 times.

We can generalize these findings: Ctr-Ctr delays corresponding to Pareto points vary much more than Sw-Ctr delays in a generic network. Indeed, Ctr-Ctr delays are  by construction between a minimum of 1-2 hops (when all the controllers are at the closest distance) and the maximum equal to the diameter of the network. The gains for the Sw-Ctr delays are lower, since  the availability of multiple controllers decreases the maximum distance to reach the master controller from a switch.
We can conclude that larger Sw-Ctr delays with respect to the minimum ones are well compensated by much smaller Ctr-Ctr delays. This highlights the relevant role of the proper design of the Ctr-Ctr plane in SDN networks. 
 
\begin{table}[!tb]
\centering
\caption{Delay reductions for the extreme Pareto-optimal placements}\label{tab:gains} 
\begin{tabular}{|c||c|c|}
\hline 
ISP & $\dfrac{\text{Sw-Ctr delay in P2}}{\text{Sw-Ctr delay in P1}}$ & 
$\dfrac{\text{Ctr-Ctr delay in P1}}{\text{Ctr-Ctr delay in P2}}$ \\
\hline
HighWinds & 6.0  & 34.8 \\
Abilene & 2.4 &  4.9 \\
York & 2.9 &  15.0 \\
\hline
\end{tabular}
\end{table}



Figs.~\ref{fig:delay-highwinds4} and \ref{fig:delay-highwinds5} show the delay tradeoff achievable for 4 and 5 controllers. Qualitatively the performance confirm our findings above for 3 controllers, even if now the absolute values of the delays for Pareto-optimal points are smaller, due to the larger number of controllers.

\ifconf
\else
 {\color{blue}

\subsection{Comparison among topologies}

 

\begin{figure}[!tb]
\centering
\includegraphics[angle=-90,width=8.2cm]{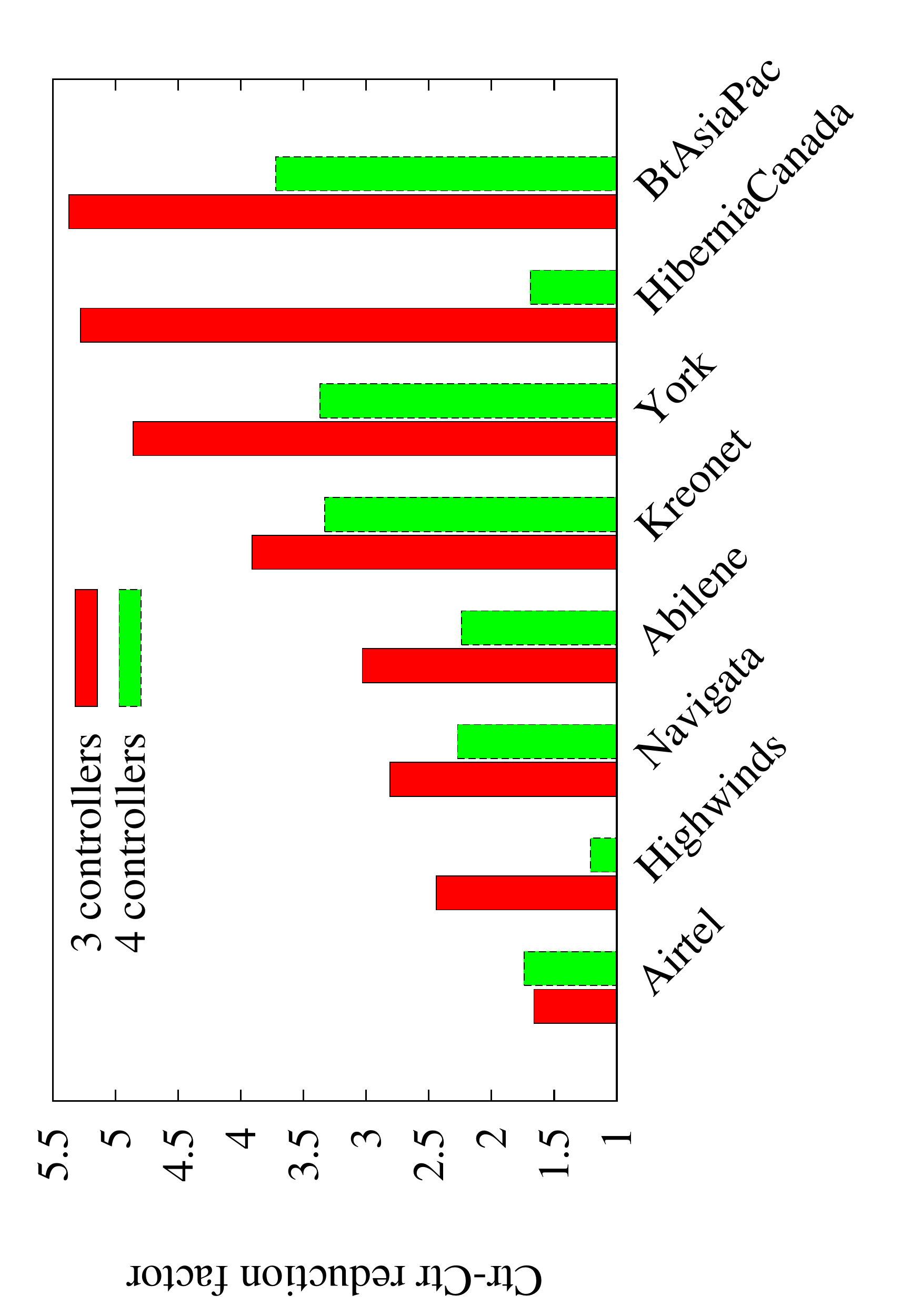}
\caption{Ctr-Ctr reduction factor when doubling the Sw-Ctr delay}
\label{fig:plot1}
\end{figure}


We now analyze the results obtained for different topologies. 
To highlight the difference with respect to a standard placement problem that minimizes the Sw-Ctr delay, we define a new metric that evaluates the reduction in Ctr-Ctr delay whenever we accept  some little increase in the Sw-Ctr delay. 
Let  $P'=(d_{sc}',d_{cc}')$ be the Pareto placement that minimizes the average delay, where $d_{sc}' $ and $d_{cc}'$ are the corresponding Sw-Ctr and Ctr-Ctr delays. 
Consider now the specific Pareto placement $P''=(d_{sc}'',d_{cc}'')$ with the minimum Ctr-Ctr delay such that the Sw-Ctr delay increases at most by a factor $2$, i.e.\
$d_{sc}''\leq2d_{sc}'$. We define the {\em Ctr-Ctr reduction factor} as the ratio ${d_{cc}'}/{d_{cc}''}$, which evaluates the relative reduction of Ctr-Ctr delay whenever we accept to double  the Sw-Ctr delay. 

Figs.~\ref{fig:plot1}  reports the  Ctr-Ctr reduction factor for different network topologies obtained for 3 and 4 controllers. The reduction is larger for 3 controllers, since the average distance between controllers is larger by construction. 
The gain for 3 controllers depends heavily on the particular topology. For the first 2 topologies, the growth in the Sw-Ctr delay is not compensated by the same reduction in Ctr-Ctr delay. Instead, for the remaining 8 topologies, the reduction in the Ctr-Ctr delay is much higher, achieving also a factor 6; in this case, increasing a little the Sw-Ctr delay has a very strong beneficial effect on the Ctr-Ctr delay.

It is interesting to note that in same cases the reduction decreases from 3 to 4 controllers (as in Highwinds and HiberniaCanada). It can be shown that this is actually due to the peculiar clustered topology of the two ISPs, 
that are similar to a single star connected to one or two nodes very far (e.g.\ in Highwinds, we have one star-like cluster in North America and very few nodes in South America and in Europe). 

}
\fi
\subsection{Reaction time for SDO and MDO models}\label{sec:rt}

\begin{figure}
\centering
\includegraphics[width=8.2cm]{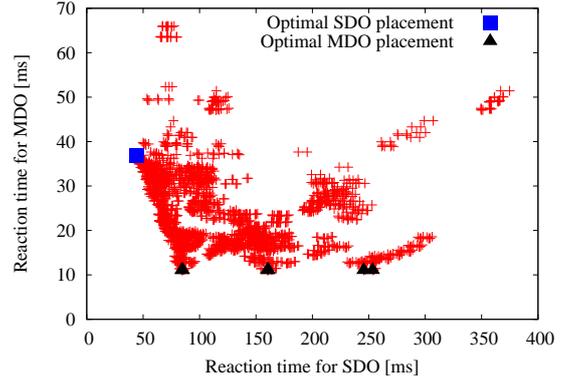}
\caption{Average reaction times in HighWinds network for all the placements. Optimal placements for the two data-ownership models are highlighted. }
\label{fig:own}
\end{figure}

\begin{figure}
\centering
\includegraphics[width=8.2cm]{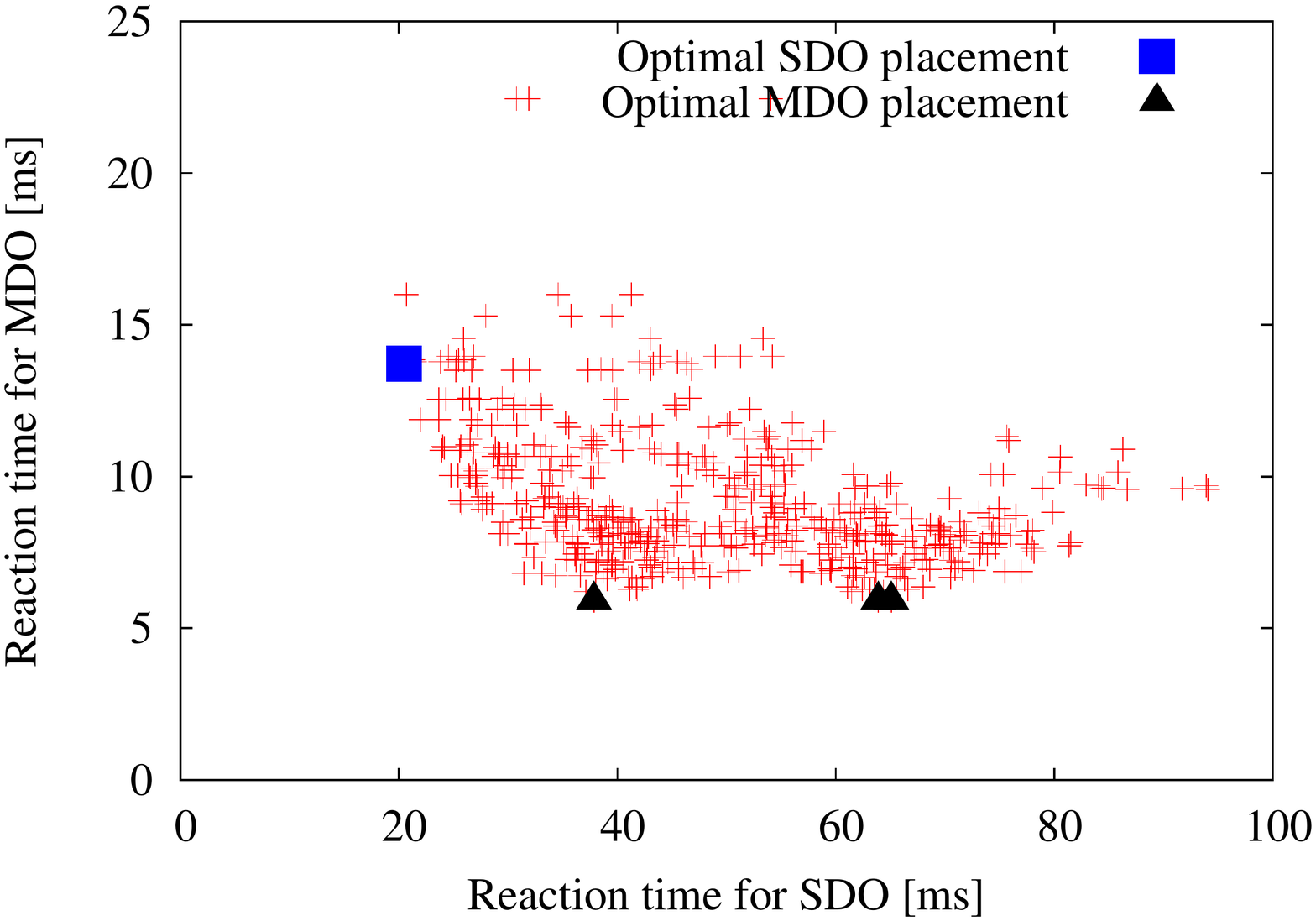}
\caption{Average reaction times in Abilene network for all the placements. Optimal placements for the two data-ownership models are highlighted. }
\label{fig:own-abi}
\end{figure}

\begin{figure}
\centering
\includegraphics[width=8.2cm]{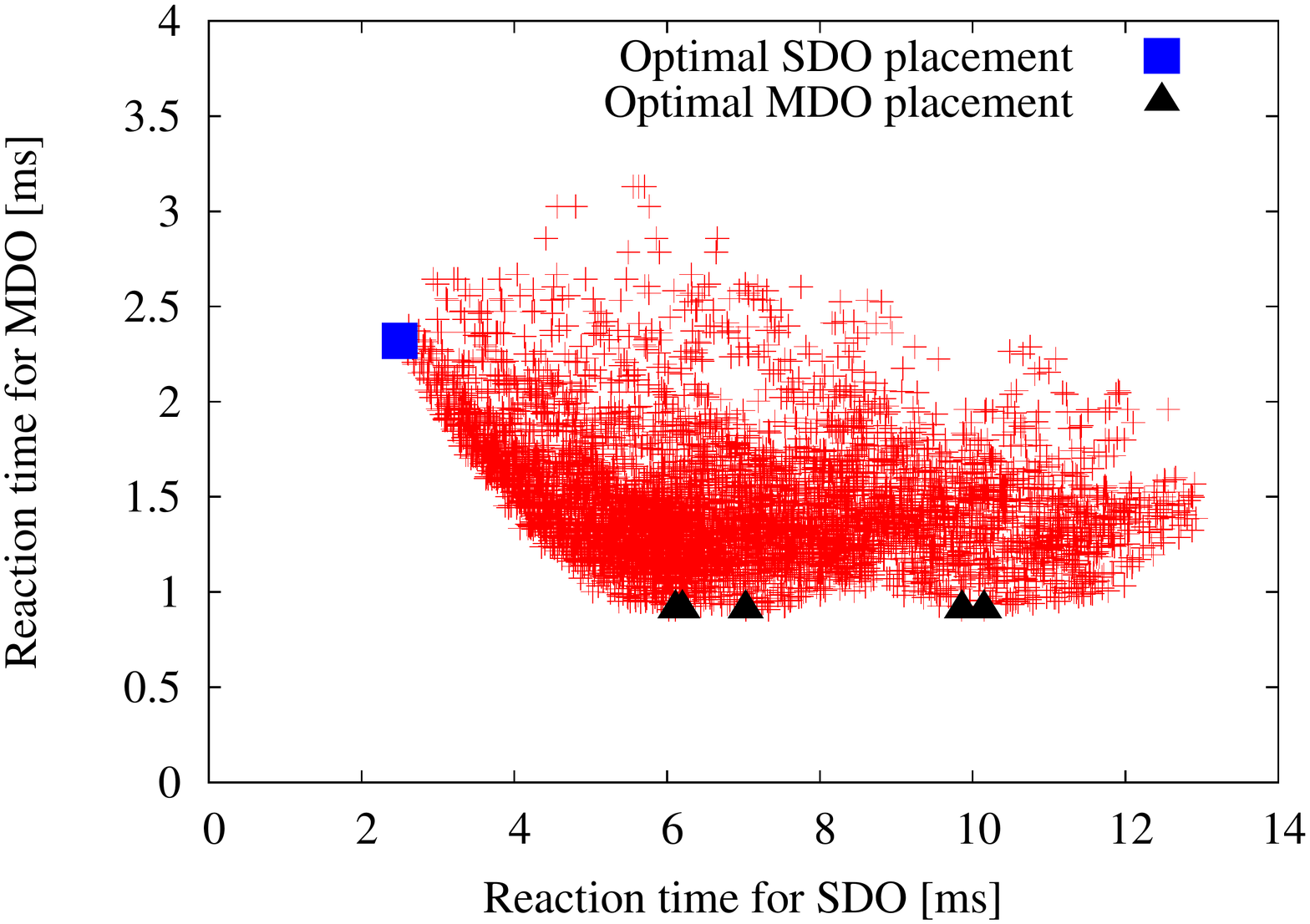}
\caption{Average reaction times in York network for all the placements. Optimal placements for the two data-ownership models are highlighted. }
\label{fig:own-york}
\end{figure}


\begin{figure}
\centering
\includegraphics[width=8.2cm]{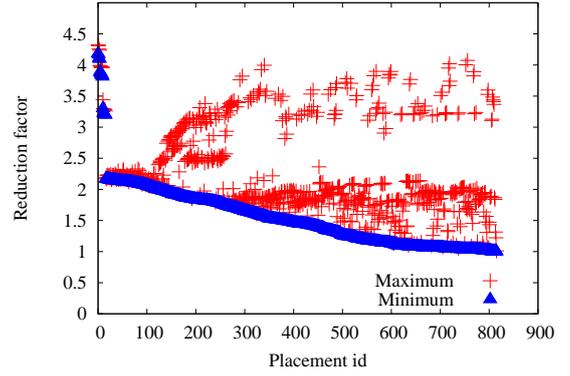}
\caption{Reaction time reduction in HighWinds network for the optimal selection of the data owner in the SDO model. }
\label{fig:benefit-30000}
\end{figure}

\begin{figure}
\centering
\includegraphics[width=8.2cm]{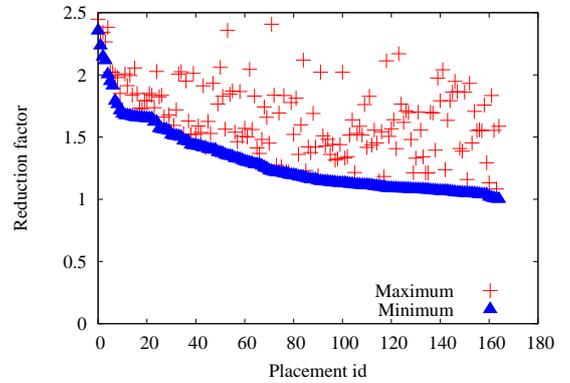}
\caption{Reaction time reduction in Abilene network for the optimal selection of the data owner in the SDO model. }
\label{fig:benefit-30000-abi}
\end{figure}

\begin{figure}
\centering
\includegraphics[width=8.2cm]{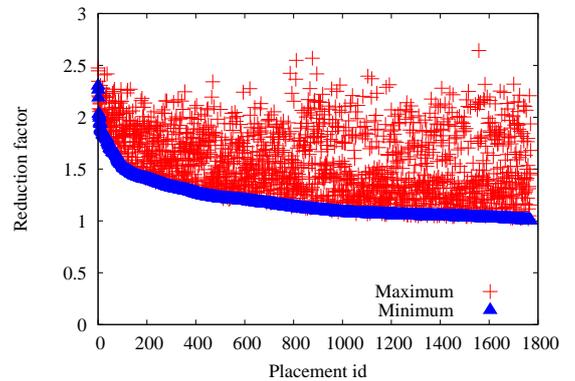}
\caption{Reaction time reduction in York network for the optimal selection of the data owner in the SDO model. }
\label{fig:benefit-30000-york}
\end{figure}

We investigate the reaction times achievable for different data-ownership
models, based on Properties~\ref{prop:1d} and \ref{prop:1s}. Given a controller placement, we study the effect of selecting
the data owner among the controllers on the perceived controller
reactivity. 
%


 In Figs.~\ref{fig:own}-\ref{fig:own-york} we report the scatter plots of the average reaction times for the SDO and the MDO models when considering all possible controllers' placements and  all possible selections for the data owner, in the case of 3 controllers.
 Each controller placement appears with 3 points aligned horizontally, one for each data owner, since the data owner selection does not affect the MDO reaction time.
In the plots we have highlighted the placements with the minimum reaction time according to the SDO and MDO models.  By construction, the minimum reaction time for the MDO is always smaller than the one for SDO model. From these results, the optimal placements are shown to be very different for the two data-ownership models and this fact motivates the need for a careful choice of the controllers placement and the owner, based on the adopted data-ownership model.

To highlight the role of the proper selection of the data owner for the SDO model, in Figs.~\ref{fig:benefit-30000}-\ref{fig:benefit-30000-york} we investigate the benefit achievable when considering the best data owner among the 3 available controllers, for the three ISPs under consideration. 
Assume that a given controller placement corresponds to three values of reaction times: $d_1$, $d_2$ and $d_3$, sorted in increasing order. The minimum {\em reduction factor} is defined as $d_2/d_1$ and the maximum reduction factor as $d_3/d_1$.
We plot the delay reduction factor due to the optimal choice of the data owner, for any possible placement.
For the sake of readability, the placements have been sorted in decreasing order of minimum reduction factor.
 Figs.~\ref{fig:benefit-30000}-\ref{fig:benefit-30000-york} show that a careful choice of the data owner in the SDO model decreases the reaction time by a factor around 2 and 4.

These results show that the selection of the data owner in the SDO model has the largest impact on the perceived performance of the controller, and can be easily optimally solved by considering all the possible $C$ cases, after having fixed the controller placement.

%% file: algorithm.tex
\ifconf
\else
{\color{blue}
\section{A evolutionary algorithm to find Pareto-optimal placement for large networks}\label{sec:algo}
In this section, we present an evolutionary algorithm, denoted as \textsc{Evo-Place}, to compute the Pareto optimal controller placements without performing an exhaustive search, as pursued in Sec.~\ref{sec:simu}, since not feasible for large networks. Our algorithm continuously optimizes both average Ctr-Ctr delay and Sw-Ctr delay through controller positions perturbation in each iteration. We run our algorithm on different network topologies with varying number of controllers, and compare the results with both the optimal Pareto frontier and the one obtained by a pure random placement algorithm. Thus, we  show that our algorithm can locate the Pareto optimal controller placements with high accuracy and with a small complexity.

\subsection{The algorithm for controller placement}
The basic idea of our algorithm is to discover new non-dominated solutions by perturbing the actual set of Pareto 
solutions for the controller placement.
Specifically, starting from a given controller placement in the network, we may get a new controller placement with better Ctr-Ctr delay by putting the controllers closer, and a new controller placement with better Sw-Ctr delay by distributing the controllers more evenly in the network. By continuously perform such perturbation, we achieve a good approximation of the Pareto frontier for the placement problem. 

We start by defining the pseudocode of a basic randomized algorithm, denoted as \textsc{Rnd-Place} and reported in Algorithm~\ref{ra}, able to find a set of Pareto solutions just using a random sampling. 
The input parameters are the number of controllers $C$, the number of nodes $N$ and the number of iterations $i_{\max}$.
We assume that function \textsc{Random-Permutation}$(N,C)$ provides the first $C$ elements of a random permutation of size $N$, with $C\in[1,N]$; its complexity is $O(C)$ thanks to the classical Knuth shuffle algorithm.
Let $P$ be the current set of all Pareto (i.e.\  non-dominated) solutions.
At each iteration, a new placement is generated (step~\ref{a:1}). Now function \textsc{Add-Prune} eventually adds $\pi$ to $P$. More precisely, if $\pi$ is dominated by any Pareto solution in $P$, then it not added to $P$ since it is not Pareto (step~\ref{a:2}). Instead, if any  current solution $p\in P$ is dominated by $\pi$, then it is removed (step~\ref{a:3}) and then $\pi$ is added as new Pareto solution (step~\ref{a:4}). 
\textsc{Add-Prune} returns true if $\pi$ was added successfully, otherwise it returns false.

The set $P$ returned by \textsc{Rnd-Place} at the end of $i_{\max}$ iterations collects all the Pareto placements found by the procedure, and corresponds to an approximation of the optimal Pareto frontier for the controller placement problem.
The randomized approach is simple but quite inefficient in terms of complexity, since it takes $O(|\Omega|\log|\Omega|)$ iterations (thanks to the well known results about the  coupon collector problem) to approach the exaustive search and find the optimal Pareto placements.

We have modified \textsc{Rnd-Place} to exploit an evolutionary approach to boost the efficiency of the algorithm.  Algorithm~\ref{ga} reports the pseudocode of our proposed \textsc{Evo-Place}. At each iteration, the algorithm selects a random placement $\pi$ (step~\ref{a:10}) and try to add to $P$, as in~\textsc{Rnd-Place}. If the addition is successful (i.e.\ $\pi$ is Pareto), then $\pi$ is perturbed (step~\ref{a:6}) and the new placement $\pi'$ is eventually added to $P$ (step~\ref{a:7}).  The loop for the perturbation ends when the newly perturbed solution cannot be added to $P$, since dominated by other solutions (steps~\ref{a:5}-\ref{a:11}).
The perturbation phase is described by  \textsc{Decrease-Ctr-Ctr-delay}, whose pseudocode is reported in Algorithm~\ref{ap1}.
 \textsc{Decrease-Ctr-Ctr-delay}  perturbs the given placement solution $\pi$ by decreasing the Ctr-Ctr delay. The intuition is to move the farthest controller closer to the others. Indeed, in steps~\ref{a:21}-\ref{a:22} the average distance is computed for each controller to all the others (actually, we omit the division by $C-1$ since useless for the following steps).
We define $d_{ij}$ as the minimum delay from node $i$ to $j$, based on the propagation delays in the network topology.
Now we choose $c'$ as the controller with the maximum average delay towards the others (step~\ref{a:23}) and find $c''$ as the closest controller to $c'$  (step~\ref{a:24}). Now we  move $c'$ one hop towards $c''$ (steps~\ref{a:25}) along the shortest path from $c'$ to $c''$; note that the check that $c''$ is at least 2 hops far from $c'$ guarantees that the movement is possible. 
As result, our approach tends to decrease the average Ctr-Ctr distance most of the times, but does not guarantee this happening always.


\begin{algorithm}[!htb]
	\caption{Random algorithm for finding Pareto controller placements}\label{ra}
	\begin{algorithmic}[1]
		\Procedure{\textsc{Rnd-Place}}{$C,N,i_{\max}$}
		\State $P=\emptyset$\Comment{Init the set of Pareto solutions}
		\For {$i=1\to i_{\max}$}\Comment{For $i_{\max}$ iterations}
			\State $\pi=$\textsc{Random-Permutation$(N,C)$}\label{a:1}
			\State \textsc{Add-Prune}$(P,\pi)$
		\EndFor 
		\State \Return $P$
		\EndProcedure
		\Procedure{\textsc{Add-Prune}}{$P,\pi$}
		\ForAll {$p\in P$}
			\If {$\pi$ is dominated by $p$}
				\State \Return false \label{a:2}\Comment{Unsuccessful addition of $\pi$}
			\EndIf
			\If{$p$ is dominated by $\pi$}
				\State $P=P\setminus \{p\}$\Comment{Remove $p$}\label{a:3}
			\EndIf
		\EndFor
		\State $P=P\cup\{\pi\}$\Comment{Add $\pi$ since not dominated}\label{a:4}
		\State \Return true \Comment{Successful addition of $\pi$}
		\EndProcedure
	\end{algorithmic}
\end{algorithm}

\begin{algorithm}[!htb]
	\caption{Evolutionary algorithm for finding Pareto controller placements}\label{ga}
	\begin{algorithmic}[1]
		\Procedure{\textsc{Evo-Place}}{$C,N,i_{\max}$}
		\State $P=\emptyset$\Comment{Init the set of Pareto solutions}
		\For {$i=1\to i_{\max}$}\Comment{For $i_{\max}$ iterations}
			\State $\pi=$\textsc{Random-Permutation$(N,C)$}\label{a:10}
			\State success$\_$flag=\textsc{Add-Prune}$(P,\pi)$
			\While{(success$\_$flag=true)}\label{a:5}
					\State $\pi'=$\textsc{Decrease-Ctr-Ctr-delay}($\pi$)\label{a:6}
					\State success$\_$flag=\textsc{Add-Prune}$(P,\pi')$\label{a:7}
					\State $\pi=\pi'$\label{a:11}
			\EndWhile
		\EndFor
		\State\Return $P$
		\EndProcedure
	\end{algorithmic}
\end{algorithm}

\begin{algorithm}[!htb]
	\caption{Perturb a given controller placement $\pi$ to decrease Ctr-Ctr delay}\label{ap1}
	\begin{algorithmic}[1]
  		%
		\Procedure{\textsc{Decrease-Ctr-Ctr-delay}}{$\pi$}
		\For {$c=1\to C$}\label{a:21}
			\State $h_c=\sum_{k\neq c} d_{\pi_c \pi_k}$\Comment{Total delay from $c$} \label{a:22}
		\EndFor 
		\State $c'=\arg \max_{c} \{h_c\}$ \Comment{Farthest controller}\label{a:23} 
		\State $c''=\displaystyle\arg\min_{c\neq c'} \{ d_{\pi_c \pi_{c'}} \}$
		\Comment{$c'$'s closest cnt.\ }\label{a:24}
		\State $n$={\bf find} first node in the shortest path from $c'$ to $c''$\label{a:25}
		\If {$n\neq \pi_{c''}$}
			\State $\pi_{c'}=n$\Comment{Move $c'$ into $n$}\label{a:26}
	       \EndIf
		\State \Return $\pi$       
  		\EndProcedure
	\end{algorithmic}
\end{algorithm}

\subsection{Results}
We have compared the performance of \textsc{Exa-Place}, \textsc{Rnd-Place} and \textsc{Evo-Place} on many different networks with varying number of controllers.
In Fig.~\ref{fig:delay-garr}, we show the results for the Garr network, a nation-wide Italian ISP, taken from~\cite{zooweb}, with 35 nodes, for the case of 3 controllers. Thus, $N=35$, $C=3$ and thus  ${35 \choose 3}=6,545$  are all the possible placements, drawn in the graph and evaluated by the exhaustive search \textsc{Exa-Place}. The corresponding Pareto points represent the optimal Pareto frontier, used as a reference for the frontiers obtained with the other algorithms.
The graphs show the sub-optimal Pareto points obtained by  \textsc{Rnd-Place} and \textsc{Evo-Place} running for $i_{\max}=50$ iterations, thus corresponding to a sampling fraction $0.8\%$  (i.e.\ $50/6,545$) of all possible solutions. From the figure, the Pareto placements computed by  \textsc{Evo-Place} appear to approximate much better the optimal ones than \textsc{Rnd-Place}, given the same number of iterations. 
\begin{figure}[!tb]
	\centering
	\includegraphics[width=1.010\columnwidth]{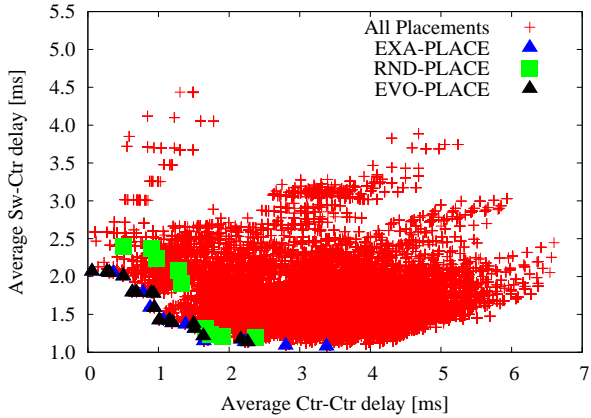}
	\caption{Optimal Pareto frontier (\textsc{Exa-Place}) and its approximations (\textsc{Rnd-Place}, \textsc{Evo-Place} with $i_{\max}=50$) for the placement of 3 controllers in Garr network} 
	\label{fig:delay-garr}
\end{figure}

In order to evaluate in a quantitative way the ``distance" between the optimal Pareto frontier computed by \textsc{Exa-Place} and the approximated ones obtained by \textsc{Rnd-Place} and \textsc{Evo-Place}, we define two error indexes, as depicted in Fig.~\ref{fig:error}, derived from classical volume based performance indexes for Pareto sets~\cite{error}: (i) the {\em average Sw-Ctr error}, computed as the average vertical distance between the optimal Pareto frontier and the approximated Pareto frontier, (ii) the   {\em average Ctr-Ctr error}, computed as the average horizontal distance between the two frontiers. 
Fig.~\ref{fig:error-garr} shows the behavior of the two errors in function of the number of iterations, in the same scenario considered in Fig.~\ref{fig:delay-garr}.
Each experiment, for a given number of iterations, was repeated multiple times to get an accurate estimation of the error.
 After 10 iterations, corresponding to a sampling ratio $10/6,545=0.15\%$, \textsc{Rnd-Place} shows an average error of 1.6~ms for the Ctr-Ctr delay whereas 0.8~ms for the Sw-Ctr delay. Given the same number of iterations, \textsc{Evo-Place} obtains a reduction in both delays by a factor 4. After 50 iterations (i.e.\ almost 1\% sampling ratio), \textsc{Evo-Place}  reaches a stable error, around 0.1 and 0.2~ms for the two delays, thus approximating quite well the optimal Pareto region. When the number of iterations is  large, the errors in the Pareto frontier obtained by \textsc{Rnd-Place} is more than twice than \textsc{Evo-Place}. This fact shows that the boost in performance due to the \textsc{Decrease-Ctr-Ctr-delay} procedure is very effective. 

\begin{figure}[!tb]
	\centering
	\includegraphics[width=6cm]{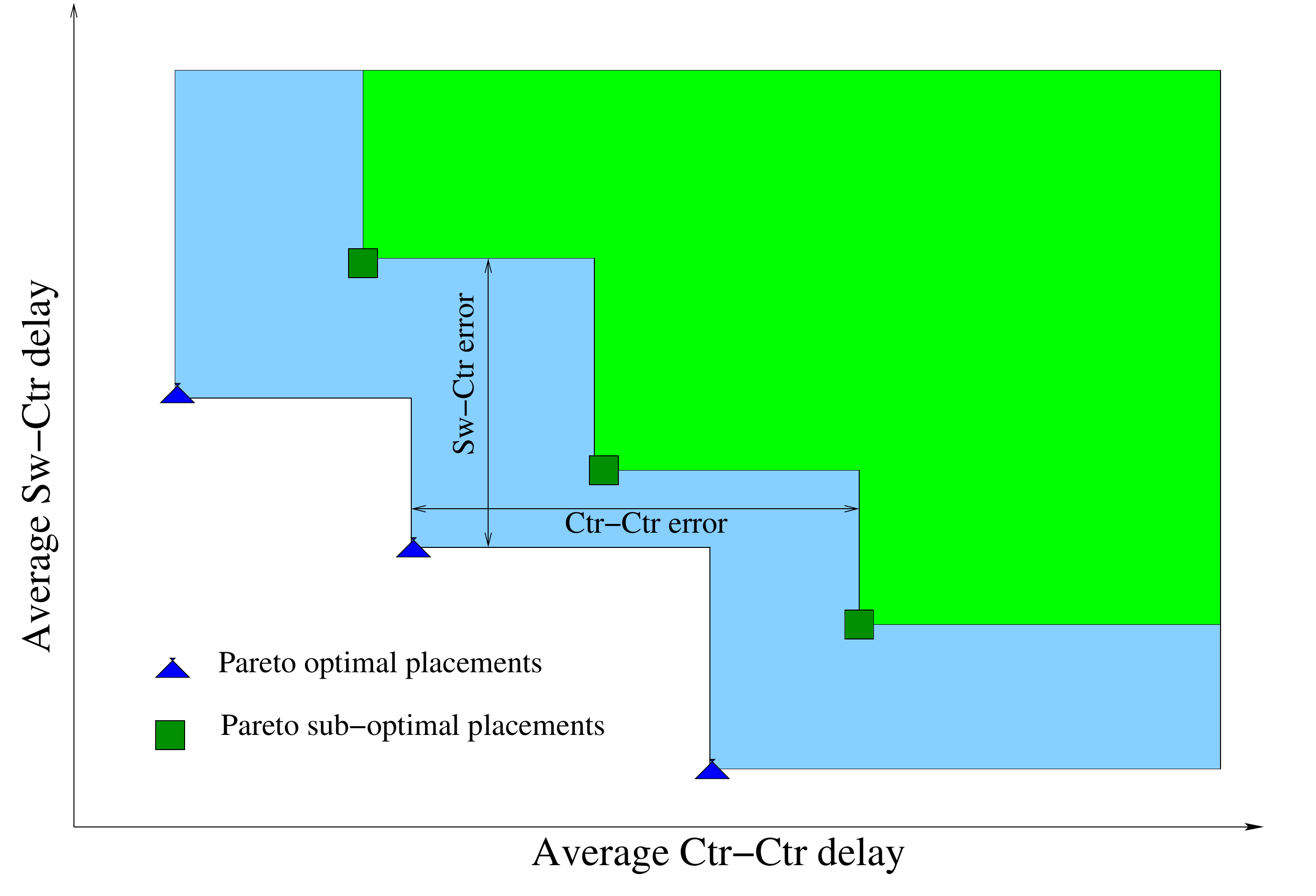}
	\caption{Definition of Ctr-Ctr error and of  Sw-Ctr error with respect to the optimal Pareto frontier} 
	\label{fig:error}
\end{figure}

\begin{figure}[!tb]
	\centering
	\includegraphics[page=1, width=1.010\columnwidth]{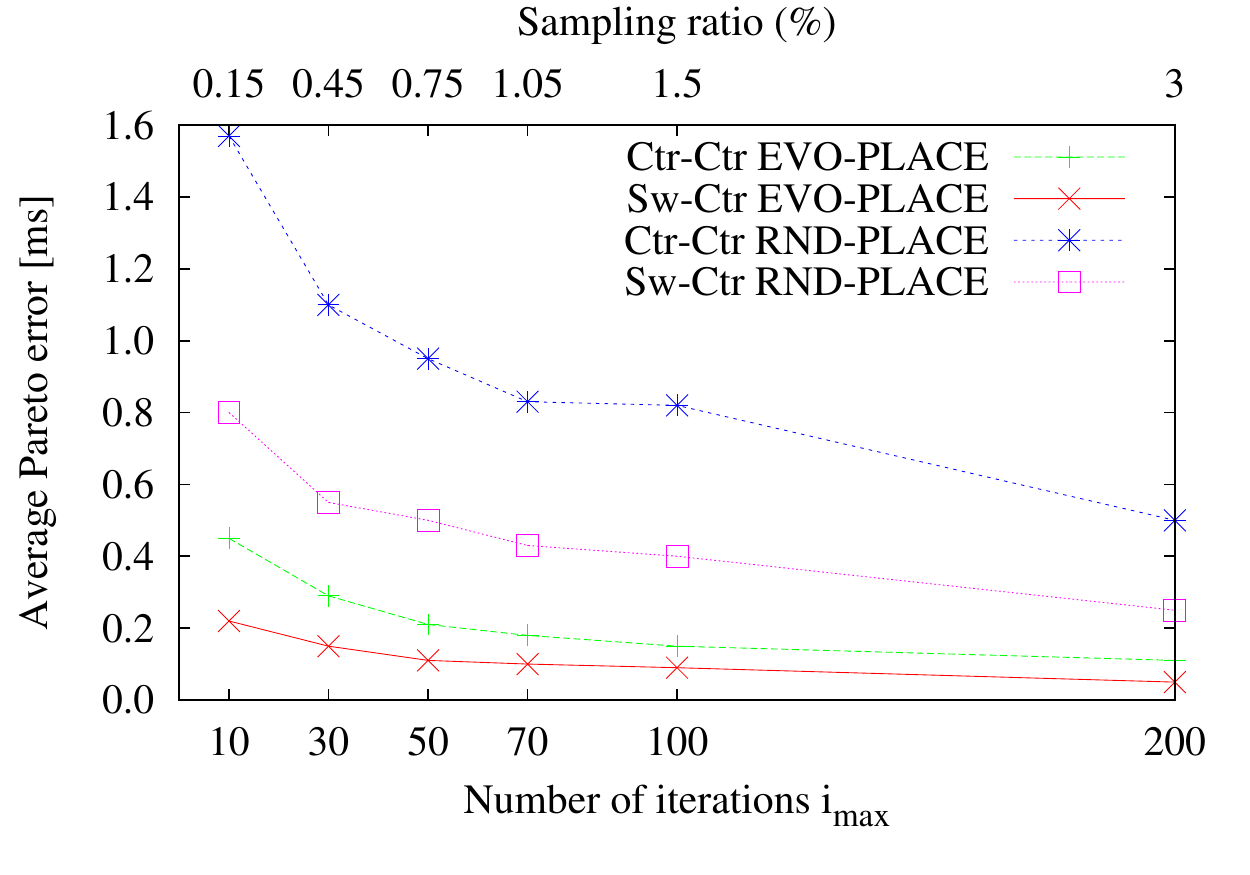}
	\caption{Pareto frontier error for Garr network} 
	\label{fig:error-garr}
\end{figure}

We extended our investigation to other larger topologies, for which the approximated approach is much faster than the exhaustive search. Figs.~\ref{fig:delay-chinanet} and \ref{fig:delay-deltacom} show the errors in the Pareto frontiers obtained for China-Telecom and ITC-Deltacom networks, respectively, taken from~\cite{zooweb}. In China-Telecom, (38 nodes) \textsc{Rnd-Place} with 10 iterations achieve errors of 1.2 and 0.6~ms, respectively, and both Sw-Ctr and Ctr-Ctr delays are  reduced by a factor 2 thanks to \textsc{Evo-Place}. This reduction factor remains quite constant also for larger number of iterations. Similarly to Garr network, around 1\% of sampling ratio \textsc{Evo-Place} tends to reach the minimum error.  

Fig.~\ref{fig:delay-deltacom} shows the errors for ITC-Deltacom network, which is a large USA ISP with 100 nodes. In this case, after only 50 iterations (corresponding to a very small sampling ratio) \textsc{Evo-Place} reaches the minimum error, which is still $2\times$ better than \textsc{Rnd-Place}. 

We have also evaluated the scenario with Colt-Telecom from~\cite{zooweb}, an Europe-wide ISP covering 149 nodes, in the case of 10 controllers. In this scenario \textsc{Exa-Place} cannot run since the total number of possible placements is larger than $10^{15}$ and thus we cannot evaluate the average errors with respect to the optimal Pareto points. We instead observe that \textsc{Evo-Place} is always outperforming \textsc{Rnd-Place} by reducing the average Sw-Crt and Ctr-Ctr delays of $0.25-1$~ms. 

In conclusion, for all the scenarios we investigated, we were able to observe a better Pareto frontier obtained by \textsc{Evo-Place} with respect to \textsc{Rnd-Place}, given the same number of iterations and thus the same computation complexity. Thus, the evolutionary approach adopted in \textsc{Evo-Place} appears efficient in finding the Pareto placements for a given network topology, especially when the network is large and an exhaustive approach is not feasible anymore.


\begin{figure}[!tb]
	\includegraphics[page=1, width=1.010\columnwidth]{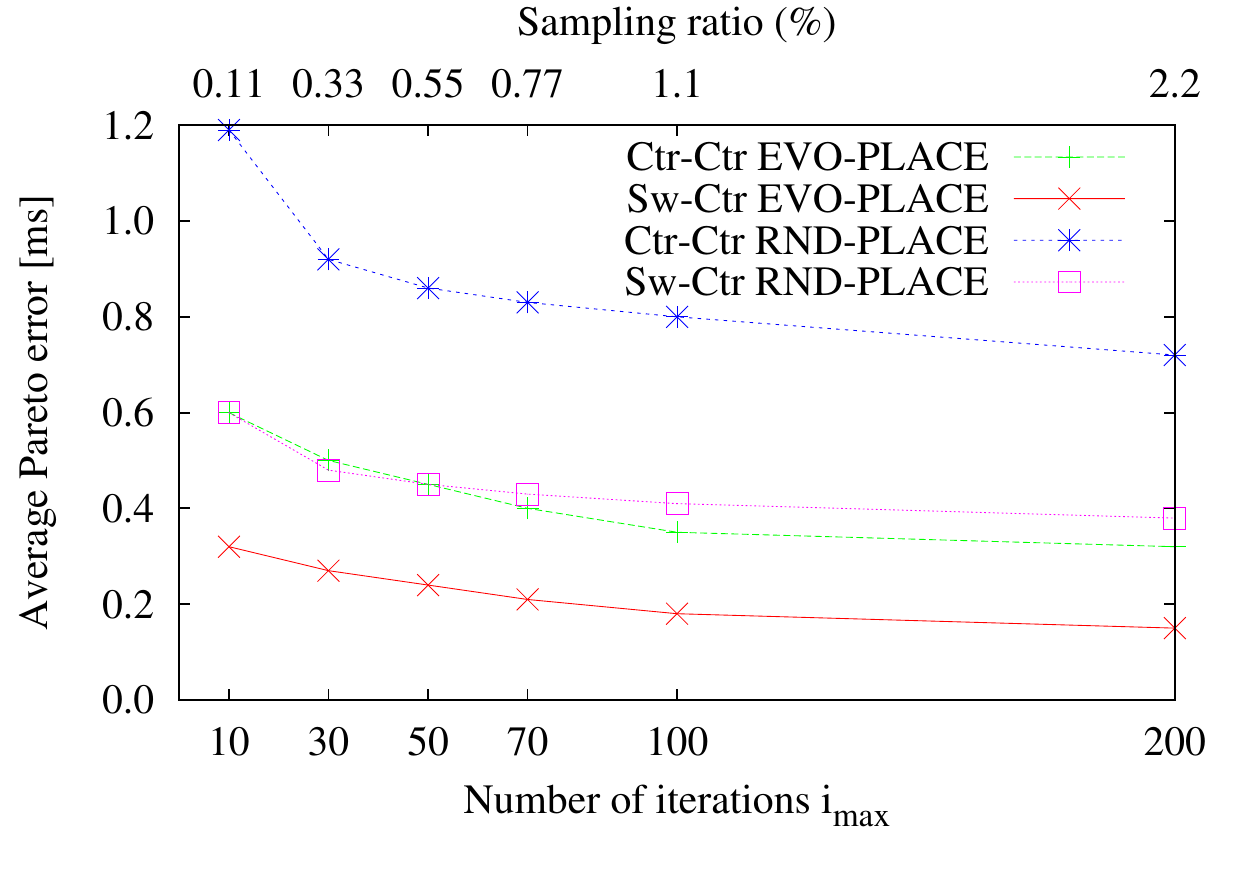}
	\caption{Pareto frontier error for Chinanet network} 
	\label{fig:delay-chinanet}
\end{figure}

\begin{figure}[!tb]
	\centering
	\includegraphics[page=1, width=1.010\columnwidth]{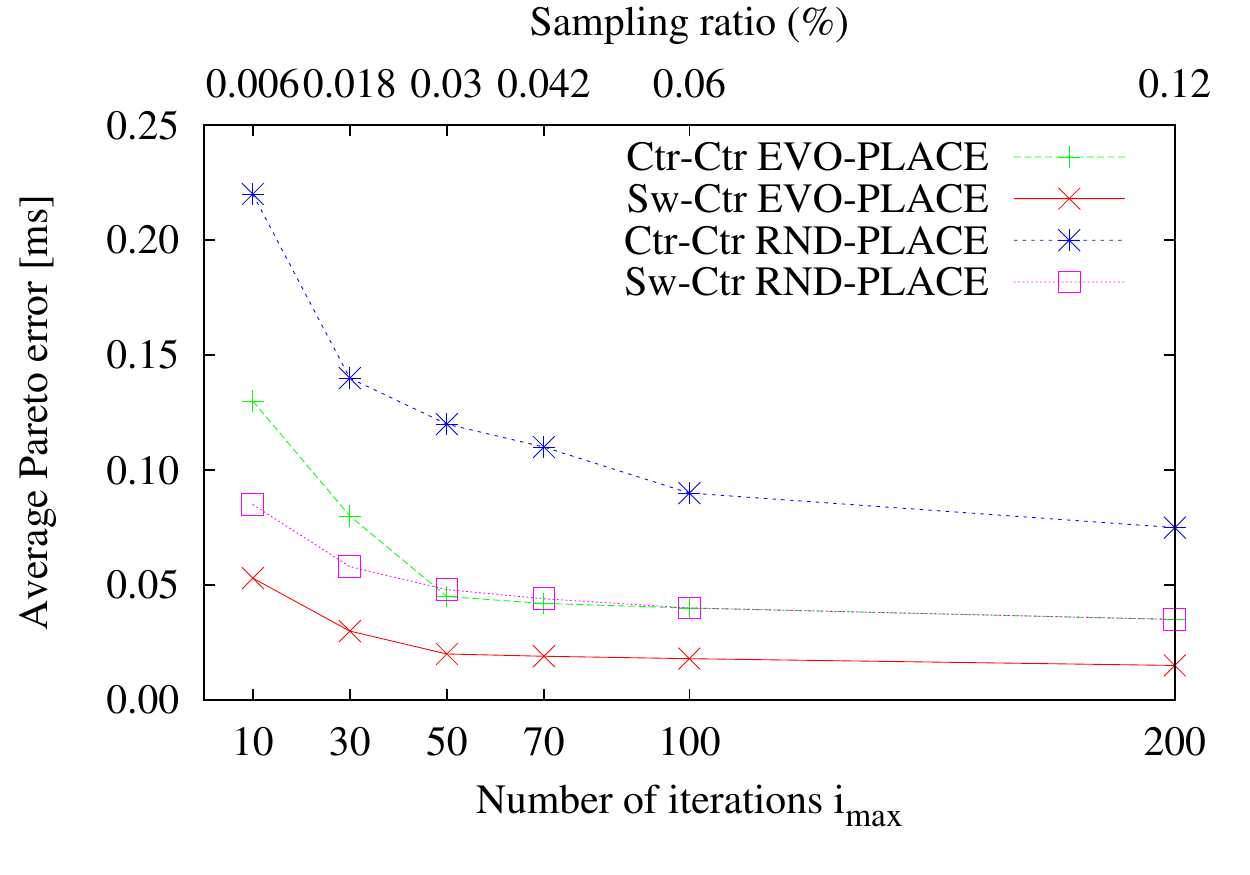}
	\caption{Pareto frontier error for Deltacom network} 
	\label{fig:delay-deltacom}
\end{figure}
}
\fi

%% file: previous.tex
\section{Related work}\label{sec:pw}
The work in~\cite{levin} emphasizes the importance of the network state consistency, and indicates that inconsistent network states degrade the performance of network applications. Thus, \cite{levin} motivates 
our work, since we devise the controller placement problem to target small Ctr-Ctr delays in the MDO model, thus improving the resilience of the network to possible state inconsistencies between controllers.

Many works address the control placement problem in SDN, but with slightly different objectives. The works~\cite{ruiz,hu} target fault tolerance, 
%
%
whereas~\cite{yao} aims at balancing the load on the controllers.
%
Both~\cite{nick12,bari} focus on the optimal controller placement by considering only the minimization of Sw-Ctr delays (average or maximum). Differently from us, they neglect completely the interaction among controllers and thus the Ctr-Ctr delays. In the case of SDO model, \cite{nick12,bari} neglects the relevant role of the data owner.
In the case of MDO, we have shown in Sec.~\ref{sec:rt} that by relaxing the minimum switch-to-controller delay target, it is possible to significantly reduce  the Ctr-Ctr delays and improve the convergence to a consistent network state.    
Similarly to our approach, \cite{hock12,poco} consider the possible Pareto optimal controllers' placements under a variety of performance and resilience factors, as controller failure tolerance, network disruption, load balancing and Ctr-Ctr delays. Their main contribution is the algorithm to find the Pareto placements, not the analysis of the structure of all the Pareto optimal solutions, as also considered in our work. 
Finally, \cite{st} provides a general mathematical framework to compute the optimal controllers' placement, under generic cost functions, but it neglects the role of Ctr-Ctr delays. 

%% file: conclusion.tex
\section{Conclusions}\label{sec:con}

We considered a distributed architecture of SDN controllers, with an in-band control plane. We investigated the performance issues related to the placement of the controllers across the network nodes. Differently from previous work, we highlighted the importance of the interaction among the controllers in the placement problem. We identified two possible models for the shared data structures: the single  and the multiple data-ownership models, which are both implemented in state-of-art controllers.
We evaluated analytically the controllers reactivity as perceived by the switches for the two 
\ifconf
\else
{\color{blue}
models, and showed the accuracy of our models in a SDWAN.}
\fi
We studied the optimal controllers placement problem taking into account all the communications in the control plane (from the switches to the controllers, and among the controllers). 
We computed the optimal Pareto frontier for some realistic ISP topologies.
\ifconf
\else
{\color{blue}
 We also proposed a new evolutionary algorithm to compute such frontier for large networks.}
 \fi
Based on our numerical results, the choice of the placement of the  specific controller with the role of data owner appears of paramount importance for the single data-ownership (SDO) model, since the reactivity of the controller depends heavily on the delay between the controllers and the leader controller. For the multiple data-ownership  (MDO) model, we studied the possible tradeoffs between controller reactivity and convergence time to reach a consistent view of the network state among the controllers.

We believe that our investigation provides a solid methodology to design the network supporting the control plane in large networks, as in the scenario of SDWANs.

%% file: main.bbl
\begin{thebibliography}{10}
\providecommand{\url}[1]{#1}
\csname url@samestyle\endcsname
\providecommand{\newblock}{\relax}
\providecommand{\bibinfo}[2]{#2}
\providecommand{\BIBentrySTDinterwordspacing}{\spaceskip=0pt\relax}
\providecommand{\BIBentryALTinterwordstretchfactor}{4}
\providecommand{\BIBentryALTinterwordspacing}{\spaceskip=\fontdimen2\font plus
\BIBentryALTinterwordstretchfactor\fontdimen3\font minus
  \fontdimen4\font\relax}
\providecommand{\BIBforeignlanguage}[2]{{%
\expandafter\ifx\csname l@#1\endcsname\relax
\typeout{** WARNING: IEEEtran.bst: No hyphenation pattern has been}%
\typeout{** loaded for the language `#1'. Using the pattern for}%
\typeout{** the default language instead.}%
\else
\language=\csname l@#1\endcsname
\fi
#2}}
\providecommand{\BIBdecl}{\relax}
\BIBdecl

\bibitem{survey15}
D.~Kreutz, F.~Ramos, P.~Esteves~Verissimo, C.~Esteve~Rothenberg,
  S.~Azodolmolky, and S.~Uhlig, ``{Software-Defined Networking}: A
  comprehensive survey,'' \emph{Proceedings of the IEEE}, vol. 103, no.~1, pp.
  14--76, Jan 2015.

\bibitem{of}
N.~McKeown, T.~Anderson, H.~Balakrishnan, G.~Parulkar, L.~Peterson, J.~Rexford,
  S.~Shenker, and J.~Turner, ``{OpenFlow}: Enabling innovation in campus
  networks,'' \emph{SIGCOMM Comput. Commun. Rev.}, vol.~38, no.~2, pp. 69--74,
  Mar. 2008.

\bibitem{cap}
E.~Brewer, ``Pushing the {CAP}: Strategies for consistency and availability,''
  \emph{Computer}, vol.~45, no.~2, pp. 23--29, Feb. 2012.

\bibitem{capnet}
A.~Panda, C.~Scott, A.~Ghodsi, T.~Koponen, and S.~Shenker, ``{CAP} for
  {Networks},'' in \emph{HotSDN}, New York, NY, USA, 2013.

\bibitem{opendaylight}
\BIBentryALTinterwordspacing
{OpenDaylight: A Linux Foundation Collaborative Project}. [Online]. Available:
  \url{http://www.opendaylight.org}
\BIBentrySTDinterwordspacing

\bibitem{onos}
P.~Berde, M.~Gerola, J.~Hart, Y.~Higuchi, M.~Kobayashi, T.~Koide, B.~Lantz,
  B.~O'Connor, P.~Radoslavov, W.~Snow, and G.~Parulkar, ``{ONOS}: Towards an
  open, distributed {SDN OS},'' in \emph{ACM HotSDN}, New York, NY, USA, 2014.

\bibitem{raft}
D.~Ongaro and J.~Ousterhout, ``In search of an understandable consensus
  algorithm,'' in \emph{Proc. USENIX Annual Technical Conference},
  Philadelphia, PA, 2014, pp. 305--320.

\bibitem{clustering}
\BIBentryALTinterwordspacing
``Opendaylight controller clustering.'' [Online]. Available:
  \url{https://wiki.opendaylight.org/view/OpenDaylight_Controller: MD-SAL:
  Architecture: Clustering}
\BIBentrySTDinterwordspacing

\bibitem{raft-code}
\BIBentryALTinterwordspacing
``Opendaylight raft consensus code review.'' [Online]. Available:
  \url{https://github.com/opendaylight/controller/tree/master/opendaylight/md-sal/sal-akka-raft/src/main/java/org/opendaylight/controller/cluster/raft}
\BIBentrySTDinterwordspacing

\bibitem{onos-roadmap}
J.~Prajakta. (2014, Dec) { ONOS Summit: ONOS Roadmap 2015 }.

\bibitem{icc16}
\BIBentryALTinterwordspacing
A.~Muqaddas, A.~Bianco, P.~Giaccone, and G.~Maier, ``Inter-controller traffic
  in {ONOS} clusters for {SDN} networks,'' in \emph{IEEE ICC}, Kuala Lumpur,
  Malaysia, May 2016. [Online]. Available:
  \url{http://www.tlc-networks.polito.it/public/faculty/paolo-giaccone/publications}
\BIBentrySTDinterwordspacing

\bibitem{fv}
R.~Sherwood, G.~Gibb, K.-K. Yap, G.~Appenzeller, M.~Casado, N.~McKeown, and
  G.~Parulkar, ``Flowvisor: A network virtualization layer,'' \emph{OpenFlow
  Switch Consortium, Tech. Rep}, pp. 1--13, 2009.

\bibitem{mininet}
\BIBentryALTinterwordspacing
``Mininet: An instant virtual network on your laptop (or other pc).'' [Online].
  Available: \url{http://mininet.org/}
\BIBentrySTDinterwordspacing

\bibitem{nick12}
B.~Heller, R.~Sherwood, and N.~McKeown, ``The controller placement problem,''
  in \emph{ACM HotSDN}, 2012, pp. 7--12.

\bibitem{zooweb}
\BIBentryALTinterwordspacing
``{The Internet Topology Zoo}.'' [Online]. Available:
  \url{http://www.topology-zoo.org/dataset.html}
\BIBentrySTDinterwordspacing

\bibitem{error}
T.~Okabe, Y.~Jin, and B.~Sendhoff, ``A critical survey of performance indices
  for multi-objective optimisation,'' in \emph{Congress on Evolutionary
  Computation (CEC)}, vol.~2, Dec. 2003, pp. 878--885.

\bibitem{levin}
D.~Levin, A.~Wundsam, B.~Heller, N.~Handigol, and A.~Feldmann, ``Logically
  centralized?: State distribution trade-offs in software defined networks,''
  in \emph{ACM HotSDN}, New York, NY, USA, 2012.

\bibitem{ruiz}
F.~J. Ros and P.~M. Ruiz, ``Five nines of southbound reliability in
  software-defined networks,'' in \emph{ACM HotSDN}, New York, NY, USA, 2014.

\bibitem{hu}
Y.~Hu, W.~Wendong, X.~Gong, X.~Que, and C.~Shiduan, ``Reliability-aware
  controller placement for software-defined networks,'' in \emph{IFIP/IEEE
  International Symposium on Integrated Network Management}, 2013, pp.
  672--675.

\bibitem{yao}
G.~Yao, J.~Bi, Y.~Li, and L.~Guo, ``On the capacitated controller placement
  problem in software defined networks,'' \emph{IEEE Communications Letters},
  vol.~18, no.~8, pp. 1339--1342, Aug 2014.

\bibitem{bari}
M.~F. Bari, A.~R. Roy, S.~R. Chowdhury, Q.~Zhang, M.~F. Zhani, R.~Ahmed, and
  R.~Boutaba, ``Dynamic controller provisioning in software defined networks.''
  in \emph{CNSM}, Zurich, Switzerland, 2013, pp. 18--25.

\bibitem{hock12}
D.~Hock, M.~Hartmann, S.~Gebert, M.~Jarschel, T.~Zinner, and P.~Tran-Gia,
  ``Pareto-optimal resilient controller placement in {SDN}-based core
  networks,'' in \emph{ITC}, Sept 2013, pp. 1--9.

\bibitem{poco}
D.~Hock, S.~Gebert, M.~Hartmann, T.~Zinner, and P.~Tran-Gia, ``{POCO}-framework
  for pareto-optimal resilient controller placement in sdn-based core
  networks,'' in \emph{IEEE Network Operations and Management Symposium}, May
  2014, pp. 1--2.

\bibitem{st}
A.~Sallahi and M.~St-Hilaire, ``Optimal model for the controller placement
  problem in software defined networks,'' \emph{IEEE Communications Letters},
  vol.~19, no.~1, pp. 30--33, Jan. 2015.

\end{thebibliography}
